\documentclass[aps,prl,
twocolumn,superscriptaddress]{revtex4-2}
\usepackage{graphicx}
\usepackage{bm}
\usepackage{amsmath}
\usepackage{amssymb}
\usepackage[colorlinks,pdfstartview=FitH]{hyperref}
\usepackage{comment}
\hypersetup{linkcolor=blue,citecolor=blue,filecolor=black,urlcolor=blue}

\renewcommand\section[1]{\emph{#1}\ ---}

\newcommand{\be}{\begin{equation}}
\newcommand{\ee}{\end{equation}}
\newcommand{\bear}{\begin{eqnarray}}
\newcommand{\eear}{\end{eqnarray}}
\newcommand{\ba}{\begin{array}}
\newcommand{\ea}{\end{array}}

\begin{document}

\author{Dan Solomon}
\affiliation{Physics Department, University of Illinois at Chicago, Chicago, 
Illinois 60607, USA}
\email{dsolom2@uic.edu, hyee@uic.edu}

\author{Ho-Ung Yee}
\affiliation{Physics Department, University of Illinois at Chicago, Chicago, 
Illinois 60607, USA}

\title{Spectral Theorems for Generalized Weyl Nodes with Impurities in a Magnetic Field}

\begin{abstract}

We prove a few spectral theorems for the density of states of a Weyl node with arbitrary topology.
We show that the density of extended states of a Weyl node with random impurity potentials remains gapless in the presence of a magnetic field. Therefore, a magnetic field precludes Anderson localization in Weyl semi-metals, when inter-node transitions are suppressed for smooth enough potentials. We also provide a rigorous quantum mechanical proof of the chiral magnetic effect for arbitrary topology of a Weyl node. 

\end{abstract}

\maketitle

\section{Introduction}
Weyl semi-metals, where certain band crossing points feature a topological character which is similar to relativistic massless fermions, provide us with an interesting bridge between condensed matter and high energy physics. While relativistic Weyl fermions \cite{weyl} obey the constraints from Lorentz invariance, e.g., the alignment of spin and momentum, the quasi-particles in Weyl semi-metals are not encumbered with such restrictions, and can possess more diverse topological character \cite{herring, armitage}. Some of the most interesting physics stemming from non-trivial topology, such as chiral anomaly \cite{Adler:1969gk,Bell:1969ts} and chiral transport phenomena \cite{Fukushima:2008xe,Kharzeev:2013ffa,Kharzeev:2015znc}, therefore find themselves naturally generalized in Weyl semi-metals; a notable example is the chiral magnetic effect \cite{Son,Burkov, Basar:2013iaa,Li}.

In condensed matter systems, impurities due to imperfect lattices are common, and how they affect the spectral and transport
properties of Weyl nodes has been a subject of great interest. It has been proposed that as the impurity strength increases there is a quantum phase transition from the semi-metal to the diffusive metal phases~\cite{Fradkin}, where the density of states at the nodal point becomes non-zero~\cite{Goswami,Ominato,Roy,Kobayashi,Sbierski,Syzranov,Slager} and a non-Fermi liquid behavior emerges~\cite{Moon}. 
However, it was also pointed out that the rare region effects could make the transition a cross-over instead~\cite{Nandkishore,Pixley3,Gurarie,Wilson,Santos} (see Refs.~\cite{Buchhold,Buchhold2,Pixley2,Santos} for current status of the issue).
For a Dirac semi-metal, i.e., two degenerate Weyl points with opposite topology, a further increase of impurity was shown to cause 3D Anderson metal-insulator transition~\cite{Pixley}. Whether there is a similar transition, when each Weyl node is separated in momentum space and inter-node couplings are suppressed for smooth potentials, is an interesting question~\cite{Altland2015,Altland2016,Makhfudz}. Obviously, the topology of the Weyl node is at the heart of these questions.

In this work, we hope to shed light on these questions by studying spectral property of a single Weyl node with impurities in a non-zero external magnetic field.
A non-trivial interplay of nodal topology and magnetic field is well-known in the physics of chiral anomaly and the chiral magnetic effect~\cite{Son:2012wh,Stephanov:2012ki}, which offers a promising way to study the fate of topology in the presence of impurities.

\section{The main spectral theorem and its consequences}
We first derive the main spectral theorem that our subsequent discussions will rely upon. We consider a system described by non-interacting quasi-particle which carries a charge $q$ for electromagnetic interactions with an external vector potential $\bm A$. 
We assume that the energy spectrum of quasi-particle is described by a 1-particle Hamiltonian $\hat H(\bm\pi)$, where $\bm\pi={\bm p}-q\bm A$ is the gauge invariant momentum. This includes the lattice models, where $\bm A$ appears as Wilson lines in the hopping matrix, and $\bm p$ is the lattice momentum. We also assume the thermodynamic limit of a large volume $V\to \infty$.

The primary object we consider is the spectral density in energy, weighted by the current operator, e.g., along the $z$ direction, 
\be
\rho_J(\varepsilon)\equiv \lim_{V\to\infty}{1\over V}{\rm Tr}\left(\hat J_z \delta(\hat H-\varepsilon)\right),
\ee
where the current operator is defined by $\hat J_z=-{\partial \hat H(\bm A)\over\partial A_z}=q{\partial \hat H\over\partial \hat p_z}$, with a constant, auxiliary value of the $z$ component of the vector potential $\bm A$, and $\delta(x)$ is the Dirac's delta function, or any smooth function that is sufficiently narrow around $x=0$ with unit area. 
Our subsequent results do not depend on the details of the shape of this function. Because of $\delta(\hat H-\varepsilon)$, only the energy eigenstates with the eigenvalues close to $\varepsilon$ contribute to $\rho_J(\varepsilon)$. This makes sure that $\rho_J(\varepsilon)$ is finite and well-defined, even in the case where the spectrum of $\hat H$ has no lower or upper bounds in energy.
Since it can be shown in general that $\langle \hat J_z\rangle=0$ for a bound state, $\rho_J(\varepsilon)$ captures only the unbound extended states in the continuum.

Our main theorem is that $\rho_J(\varepsilon)$ is a constant, independent of $\varepsilon$:
\be
{\rm Theorem\,\, I :}\quad \rho_J(\varepsilon)=C_J\,\, (\rm constant).
\ee
We prove by showing that ${\partial \over\partial\varepsilon}\rho_J(\varepsilon)=0$, under reasonable assumption of smoothness and regularity of the density of states, 
\be
\rho(\varepsilon)\equiv \lim_{V\to\infty}{1\over V}{\rm Tr}\left(\delta(\hat H-\varepsilon)\right).
\ee
We have $
{\partial \over\partial\varepsilon}\rho_J(\varepsilon)=-\lim_{V\to\infty}{1\over V}{\rm Tr}\left(\hat J_z \delta'(\hat H-\varepsilon)\right)
$, and using the definition of $\hat J_z=-{\partial \hat H\over \partial A_z}$, we arrive at ${\partial \over\partial\varepsilon}\rho_J(\varepsilon)={\partial\over\partial A_z}\rho(\varepsilon,\bm A)$, where the dependence on $\bm A$ is implicit in the Hamiltonian $\hat H(\bm{\hat\pi})$. We now utilize the following observation: for any eigenstate $\psi(\bm x)$ of energy $\varepsilon$ of the Hamiltonian $\hat{H}(\bm{\hat\pi})$, the state  ${{e}^{i{\bm\alpha\cdot \bm x)}/{\hbar }}}\psi(\bm x)$ with a constant $\bm\alpha$ is an eigenstate of the Hamiltonian $\hat{H}(\bm{\hat\pi}-\bm\alpha)$ with the same energy. This corresponds to a shift of $\bm A\to \bm A+\bm\alpha/q$, which implies that the spectrum of $\hat H$, and hence the spectral density $\rho(\varepsilon)$, remains invariant under a constant shift of $\bm A$. This proves that ${\partial \over\partial\varepsilon}\rho_J(\varepsilon)={\partial\over\partial A_z}\rho(\varepsilon,\bm A)=0$.

Although the theorem and its proof is simple, its implications are rich.
Suppose that the quasi-particles are fermions, and the system has a chemical potential $\mu$ at zero temperature, and all eigenstates of energy $\varepsilon\le \mu$ are occupied. 
The current density, i.e., the current per unit volume, of the system is then given by $j_z=\int_{-\infty}^\mu \rho_J(\varepsilon)d\varepsilon$, which can be shown by 
\be
j_z={1\over V}\sum_\alpha \langle \psi_\alpha|\hat J_z|\psi_\alpha\rangle \Theta(\mu-\varepsilon_\alpha),
\ee
where $|\psi_\alpha\rangle$ are the energy eigenstates with energy $\varepsilon_\alpha$, labeled by $\alpha$, and the fact that $\rho_J(\varepsilon)$ can be written as 
\be
\rho_J(\varepsilon)={1\over V}\sum_\alpha \langle \psi_\alpha|\hat J_z|\psi_\alpha\rangle \delta(\varepsilon_\alpha-\varepsilon).
\ee
The theorem then implies that $j_z=C_J\int^\mu_{-\infty}d\varepsilon$, which is infinite, unless $C_J=0$. 

Although the appearance of infinity for $j_z$ looks troublesome, it is in fact expected by the following reason. Suppose the spectrum of $\hat H$ has an upper or lower bound. If $\varepsilon$ is outside of these bounds, there is no state to contribute to $\rho_J(\varepsilon)$, and $\rho_J(\varepsilon)=0$ for such $\varepsilon$. The theorem then implies that $\rho_J(\varepsilon)=0$ for all $\varepsilon$, i.e., $C_J=0$. Therefore, a non vanishing $C_J$ is possible only if the spectrum has no upper or lower bounds. An example we will discuss in more detail later is provided by the spectrum of a relativistic Weyl fermion. We will also consider an exception to the statement, when the density of states $\rho(\varepsilon)$ is allowed to possess a non-integrable singularity in $\varepsilon$.
The necessity of unbounded spectrum for non vanishing $\rho_J(\varepsilon)$ can also be understood by considering $N(\varepsilon)\equiv\lim_{V\to\infty}{1\over V}{\rm Tr}\left(\Theta(\varepsilon-\hat H)\right)$, where $\Theta(x)$ is the step function, which counts the number of states whose energy is less than $\varepsilon$. The similar steps as above shows $\rho_J(\varepsilon)={\partial\over\partial A_z}N(\varepsilon)=0$, since the energy spectrum, and hence $N(\varepsilon)$, do not depend on $A_z$. What saves a non vanishing $\rho_J(\varepsilon)$ for the unbounded spectrum is that $N(\varepsilon)$ is infinite and ill-defined.

The above discussion leads to the following corollary of the theorem, which is an alternative proof of 
the Bloch's theorem~\cite{Bohm,Yamamoto:2015fxa}: 
{\it The system with a lower bound in spectrum has no persistent current in any thermodynamic ensemble}. The proof relies on the fact that any thermodynamic ensemble is defined by the occupation number of 1-particle states, which depends only on its energy eigenvalue $\varepsilon$. For example, a grand canonical ensemble of temperature $T$ and chemical potential $\mu$ gives the current density,  
\be
j_z={1\over V}\sum_\alpha \langle\psi_\alpha|\hat J_z|\psi_\alpha\rangle n_F(\varepsilon_\alpha)=\int_{-\infty}^{\infty} n_F(\varepsilon)\rho_J(\varepsilon)d\varepsilon,
\ee
where $n_F(x)=(1+e^{(x-\mu)/(k_BT)})^{-1}$.
A lower bound on the spectrum implies $C_J=0$, i.e., $\rho_J(\varepsilon)=0$ for all $\varepsilon$.

When the spectrum has no bounds, and $C_J\neq 0$, a proper way to proceed is to first define the ground state (or vacuum state) of the system, e.g., the state at $T=\mu=0$ where all 1-particle states of $\varepsilon\le 0$ are occupied, and the current density $j_z$ is measured with respect to the value of the ground state. This gives, for example, in the grand canonical ensemble, $j_z=\int^\infty_{-\infty} (n_F(\varepsilon)-\Theta(-\varepsilon))\rho_J(\varepsilon)d\varepsilon=C_J\mu$, which is linear in $\mu$, and more interestingly, independent of $T$. 
This leads to the corollary: {\it A persistent current in the grand canonical ensemble, if it is non vanishing, is strictly linear in $\mu$ and independent of $T$.}
The theorem and its consequences do not assume any details of the Hamiltonian or external conditions. 

The theorem implies a strong robustness of $\rho_J(\varepsilon)$ under any reasonable perturbation.
Since $\rho_J(\varepsilon)=C_J$ is independent of $\varepsilon$, the constant $C_J$ is a property of the 1-particle states with arbitrarily large energy $\varepsilon$, and therefore its value is robust under any perturbation that could affect only the states with a finite $\varepsilon$. 
This is a non-trivial conclusion, since a perturbation in general may affect the states of small $\varepsilon$ in significant ways, e.g., the density of states and the current expectation values will be modified substantially for small $\varepsilon$. Yet, we find that $\rho_J(\varepsilon)$ should remain the same for {\it all} energy $\varepsilon$, since it should be the same for $\varepsilon\to \infty$. We will establish this robustness more rigorously in the following discussions.

 \section{A toy example}
To illustrate the theorem and its consequences in a simple example, we consider a 1-dimensional system described by the Hamiltonian 
\be
\hat H=f(\hat p_z)+V_0(z)\equiv \hat H_0+V_0(z),
\ee
where $f(\hat p_z)$ is an arbitrary function on the momentum operator $\hat p_z=-i\hbar\partial_z$, and $V_0(z)$ is a potential.
Let us ignore the potential for a moment, and consider the spectrum of $\hat H_0$, which is parametrized by the momentum eigenvalue $p_z$ of the eigenstate $\psi_{p_z}(z)=e^{-ip_z z/\hbar}$, i.e., $\varepsilon(p_z)=f(p_z)$. The density of states in momentum space is ${L\over 2\pi\hbar}dp_z$, where $L$ is the size of the system. Then, the density of states in energy is $\rho(\varepsilon)\equiv {1\over L}{\rm Tr}\left(\delta(\hat H_0-\varepsilon)\right)={1\over 2\pi\hbar}\int dp_z \delta(f(p_z)-\varepsilon)={1\over 2\pi\hbar}\sum_\alpha {1\over |f'(p_\alpha)|}$, where $p_\alpha$ are the solutions of $f(p_\alpha)=\varepsilon$. The current operator is $\hat J_z=qf'(\hat p_z)$, and our current weighted spectral density is $\rho_J(\varepsilon)={1\over L}{\rm Tr}\left(\hat J_z\delta(\hat H_0-\varepsilon)\right)={q\over 2\pi\hbar}\sum_\alpha {f'(p_\alpha)\over|f'(p_\alpha)|}$. Although $\rho(\varepsilon)$ depends on both $\varepsilon$ and the shape of $f(p_z)$, it is easy to see that $\rho_J(\varepsilon)$, which counts the number of signed crossings of the curve $f(p_z)$ at energy $\varepsilon$, is 
constant in $\varepsilon$. It coincides with a well-known topological index of 1-dimensional Hamiltonian.

When the potential is present, $p_z$ is no longer a good quantum number. The problem is still solvable exactly if $\hat H_0=\hat p_z$. The energy eigenstates are $\psi_\varepsilon(z)=e^{{i\over\hbar}\left(p_z z-\int^z_{z_0} V_0(z')dz'\right)}$, with the spectrum $\varepsilon(p_z)=p_z$. The current operator is an identity $\hat J_z=q{\partial\hat H\over \partial \hat p_z}=q$, and we find $\rho_J(\varepsilon)={q\over 2\pi\hbar}$, which is constant, independent of the potential $V_0(z)$. For a general $f(p_z)$ without bounds, i.e., $|f(p_z)|\to\infty$ as $p_z\to\pm\infty$, we can find the eigenstates of large energy in the eikonal approximation, $\psi_{p_z}(z)\sim e^{{i\over\hbar}\left(p_z z+\phi(z)\right)}$, where $\phi(z)=-{1\over f'(p_z)}\int^z_{z_0}V_0(z')dz'$, and the spectrum is $\varepsilon(p_z)=f(p_z)$ which defines $p_z$ given $\varepsilon$. The small parameter for the approximation is $V_0(z)/f(p_z)\to 0$ as $p_z\to\infty$, and
the approximation becomes exact in $\varepsilon\to\infty$ limit. The current operator is $\hat J=qf'(\hat p_z)$, with the expectation value $\langle \psi_{p_z}|\hat J_z|\psi_{p_z}\rangle=qf'(p_z)$, up to the same corrections of $V_0(z)/f(p_z)\to 0$ in the eikonal limit. We then find that $\rho_J(\varepsilon)$ is given by the same expression, ${q\over 2\pi\hbar}\sum_\alpha {f'(p_\alpha)\over|f'(p_\alpha)|}$, as $\varepsilon\to\infty$, and hence for all $\varepsilon$ according to our theorem, irrespective of the potential $V_0(z)$. 

As a corollary, our result proves {\it the absence of Anderson localization in 1-dimensional systems with a kinetic operator $f(p_z)\sim p_z^{2n+1}$ as $|p_z|\to\infty$}, since we have $\rho_J(\varepsilon)={q\over 2\pi\hbar}\neq 0$, which implies the absence of a gap in the spectrum of extended states with random potentials.

\section{Discussion on a counter example}
In this short digression, we discuss the following counter example in 1-dimensions: $\hat H=\varepsilon_0 \tanh(\hat p_z)$, where $\varepsilon_0$ is a constant. For the interval $|\varepsilon|<\varepsilon_0$, the spectrum is a monotonically increasing function on $p_z$, and we have $\rho_J(\varepsilon)={q\over 2\pi\hbar}$, while outside of the interval, no states exist and $\rho_J(\varepsilon)=0$. What causes the discontinuity in $\rho_J(\varepsilon)$ across $\varepsilon=\pm\varepsilon_0$ is the infinite number of states accumulated around the energy $\pm\varepsilon_0$. The density of states is $\rho(\varepsilon)={1\over 2\pi\hbar \tanh'(\tanh^{-1}(\varepsilon/\varepsilon_0))}={1\over 2\pi\hbar (1-(\varepsilon/\varepsilon_0)^2)}$, which is non integrable at $\pm\varepsilon_0$. The theorem applies only when the density of states is reasonably smooth and integrable in the energy interval of interests.

In the following, we discuss our main application of the Theorem I; we will compute the constant $C_J=\rho_J(\varepsilon)$ for a Weyl node of arbitrary topology in the presence of a magnetic field and impurities.

\section{ A clean Weyl node in a magnetic field}
Let us first ignore impurities for a while, and compute $C_J$ for a clean Weyl node in the presence of a magnetic field. Our main technique is based on the following observation: since $\rho_J(\varepsilon)=C_J$ is independent of $\varepsilon$, we can express $C_J$ by
\be \label{eqn:hteqcj}
C_J={1\over\sqrt{\pi} M}\int_{-\infty}^\infty d\varepsilon\, \rho_J(\varepsilon)e^{-\frac{\varepsilon^2}{ M^2}}={1\over\sqrt{\pi}M}{1\over V}{\rm Tr}\left(\hat J_z e^{-\frac{\hat H^2}{M^2}}\right),
\ee
for any $M>0$. The $M\to 0$ limit reproduces the definition of $\rho_J(\varepsilon)$, while $M\to\infty$ limit is similar to the common strategy in the proof of the index theorems.  

Before considering the most general form of a Weyl node, let us 
consider a concrete example of the simplest (relativistic) Weyl Hamiltonian in the presence of a constant magnetic field $\bm {B}=B\bm {\hat{z}}$, where $\hat{\bm z}$ is the unit vector in the $z$-direction,
\be \label{eqn:stdH}
\hat{H}=\bm\sigma\cdot\hat{\bm \pi}={{\sigma }_{x}}{{\hat{p}}_{x}}+{{\sigma }_{y}}\left( {{{\hat{p}}}_{y}}-qBx \right)+{{\sigma }_{z}}{{\hat{p}}_{z}},
\ee
where ${{\sigma }_{i}}$ are the Pauli matrices, and we work in the Landau gauge $\bm {A}=\left( 0,Bx,0 \right)$. In this case we are able to compute the right-hand side of Eq.(\ref{eqn:hteqcj}) exactly for any $M$ to find that it is indeed $M$-independent, which demonstrates the validity of our Theorem I.
 We have ${{\hat{H}}^{2}}={{\hat{H}}_{0}}+{{\hat{H}}_{I}}$, where ${{\hat{H}}_{0}}={{\left( {{{\hat{p}}}_{x}} \right)}^{2}}+{{\left( {{{\hat{p}}}_{y}}-qBx \right)}^{2}}+{{\left( {{{\hat{p}}}_{z}} \right)}^{2}}$
 and ${{\hat{H}}_{I}}=-\hbar qB{{\sigma }_{z}}$. Since ${{\hat{J}}_{z}}=q{{\sigma }_{z}}$, the Eq.(\ref{eqn:hteqcj}) becomes,
\be \label{eqn:stdCj}
{{C}_{J}}=\frac{q}{\sqrt{\pi }M}\frac{1}{V}{\rm Tr}\left( {{\sigma }_{z}}{{e}^{-\frac{1}{{{M}^{2}}}\left( {{{\hat{H}}}_{0}}-q\hbar B{{\sigma }_{z}} \right)}} \right).
\ee
To evaluate this first note that $\hbar qB{{\sigma }_{z}}$ commutes with ${{\hat{H}}_{0}}$ and that  
$ {{e}^{\frac{q\hbar B{{\sigma }_{z}}}{{{M}^{2}}}}}=\cosh \left( \frac{q\hbar B}{{{M}^{2}}} \right)+{{\sigma }_{z}}\sinh \left( \frac{q\hbar B}{{{M}^{2}}} \right)$.
Taking the spin trace we obtain
\be
{{C}_{J}}=\frac{q}{\sqrt{\pi }M}\frac{2\sinh \left( \frac{q\hbar B}{{{M}^{2}}} \right)}{V}{\rm Tr}\left( {{e}^{-\frac{{{{\hat{H}}}_{0}}}{{{M}^{2}}}}} \right).\label{eq11}
\ee
To evaluate the remaining trace in position-momentum space, we note that ${\hat{H}}_{0}$ is the classical Landau-level problem for a relativistic scalar particle.  The eigenfunctions are given by $\left| {{\psi }_{n,p_y,p_z}} \right\rangle ={{e}^{\frac{i}{\hbar }\left( {{p}_{y}}y+{{p}_{z}}z \right)}}\left| {{\phi }_{n,{{p}_{y}}}} \right\rangle$, where $\left| {{\phi }_{n,{{p}_{y}}}} \right\rangle$ are the 2D Landau wave functions, and the eigenvalues are ${{E}_{n,p_z}}=2q\hbar B\left( n+1/2\  \right)+p_{z}^{2}$, with the degeneracy per unit transverse area $g_\perp={qB\over 2\pi\hbar}$. From these we obtain
\be
{\rm Tr}\left({{e}^{-\frac{{{\hat{H}}}_{0}}{{M}^{2}}\;}} \right)=\frac{qVB}{(2\pi \hbar)^2 }\int_{-\infty }^{+\infty }dp_z{e}^{-\frac{p_{z}^{2}}{{M}^{2}}}\sum\limits_{n=0}^{\infty }{{{e}^{-\frac{2q\hbar B}{{{M}^{2}}}\left( n+\frac{1}{2} \right)}}},
\ee
which is evaluated to be $\frac{qVB\sqrt{\pi }M}{{{\left( 2\pi \hbar  \right)}^{2}}}\frac{1}{2\sinh \left( q\hbar B/{{M}^{2}} \right)}$. Using this in Eq.(\ref{eq11}), we indeed find the $M$-independent result,
\be
{{C}_{J}}=q^2B/{{\left( 2\pi \hbar  \right)}^{2}}.
\ee

We now consider the most general form of a Weyl node in the presence of a constant magnetic field $\bm B=B\hat{\bm z}$,
\be \label{eqn:exham}
\hat H= \bm F({\bm{\hat\pi}})\cdot\bm\sigma,
\ee
where $\bm F(\bm{\hat\pi})$ is an arbitrary vector valued function in $\bm{\hat\pi}$.
We will use Eq.(\ref{eqn:hteqcj}) in $M\to\infty$ limit to compute $C_J$, since it is not possible to solve the eigenvalue problem exactly. 
 Using $[\hat \pi_x,\hat \pi_y]=i\hbar q B$ and $\sigma^i\sigma^j=\delta^{ij}+i\epsilon^{ijk}\sigma^k$, we have
$
\hat H^2=\bm F^2(\bm{\hat\pi})-\hbar qB\left({\partial \bm F\over\partial \pi_x}\times {\partial \bm F\over\partial\pi_y}\right)\cdot\bm\sigma\equiv \hat H_0+\hat H_I
$. In obtaining this we neglected the terms arising from position-momentum commutators, since those terms possess less powers of momentum than the above leading term. As we will see, a finite contribution in $M\to\infty$ limit comes only from $\bm p\to\infty$ region, and the terms with less powers of momentum become irrelevant compared to the leading term.
Following similar steps in the proofs of index theorems, we invoke the heat kernel expansion
\be \label{eqn:htkerExp}
e^{-{\hat H^2\over M^2}}=e^{-{\hat H_0\over M^2}}\left(1-{1\over M^2}\int_0^1 dt\, e^{t{\hat H_0\over M^2}} 
\hat H_I e^{-t {\hat H_0\over M^2}}\right)+{\cal O}\left({1\over M^4}\right).
\ee
Since $\hat J_z=q\left({\partial\bm F\over \partial \hat p_z}\right)\cdot\bm\sigma$, the first non vanishing trace in spin space appears with the second term in the expansion, and we find
\be
C_J={2\hbar q^2 B\over \sqrt{\pi} M^3}{1\over V}{\rm Tr}\left(e^{-{\bm F^2(\bm{\hat \pi)}\over M^2}}
\left({\partial \bm F\over\partial \hat \pi_x}\times {\partial \bm F\over\partial \hat \pi_y}\right)\cdot{\partial\bm F\over\partial \hat p_z}\right),
\ee
up to sub-leading corrections from position-momentum commutators. We then evaluate the remaining trace in momentum basis, and
in large $M$ limit the trace is dominated by large momentum $\bm p$, for which $\bm\pi$ can be replaced by $\bm p$ up to sub-leading corrections, and we arrive at
\be
C_J={2\hbar q^2 B\over \sqrt{\pi} M^3}\int{d^3 \bm p\over (2\pi\hbar)^3}\,e^{-{\bm F^2(\bm p)\over M^2}}
\left({\partial \bm F\over\partial p_x}\times {\partial \bm F\over\partial p_y}\right)\cdot{\partial\bm F\over\partial p_z}.
\ee
Noting that $\left({\partial \bm F\over\partial p_x}\times {\partial \bm F\over\partial p_y}\right)\cdot{\partial\bm F\over\partial p_z}$ is the signed Jacobian for the map $\bm p\to\bm F(\bm p)$, which allows us to perform a change of variables $\bm p\to\bm F(\bm p)$ in the above integration, we find 
\be
C_J={2\hbar q^2 B\over \sqrt{\pi} M^3}N_{\bm F}\int {d^3 \bm F\over (2\pi\hbar)^3}\,e^{-{\bm F^2/M^2}}={q^2 B\over (2\pi\hbar)^2}N_{\bm F}, \label{eq18}
\ee 
where $N_{\bm F}\in \pi_2(S^2)={\bf Z}$ is the signed winding number of the map $\bm p\to\bm F(\bm p)$ at asymptotic infinity, $S_\infty^2\to S_\infty^2$, which defines the topology of the Weyl node with the Hamiltonian $\hat H=\bm F(\bm {\hat p})\cdot\bm\sigma$~\cite{Yee:2019rot}. The integer $N_{\bm F}$ is essentially the number of times the $\bm F$-space at infinity is covered in the map $\bm p\to\bm F(\bm p)$. As an example, $N_{\bf F}=N$ for $F^\pm=(p^\pm)^N$ and $F_z=p_z$, where $F^\pm=F_x\pm i F_y$ and $p^\pm=p_x\pm i p_y$.
Only the asymptotic region of $S_\infty^2$ at infinity in both $\bm p$ and $\bm F$ spaces matters, since the surviving contribution in $M\to\infty$ limit comes only from this region as seen in the $\bm F$ integration in Eq.(\ref{eq18}).

We can interpret the above result in the following interesting way. In the Landau gauge the eigenstates of $\hat H$ can be written as $\psi(\bm x)=\psi_{p_z}^{(n)}\left(x-p_y/(qB)\right)e^{{i\over\hbar}(p_y y+p_z z)}$, with good quantum numbers $(p_y,p_z)$. The eigenvalue equation takes the form $
\hat H_{p_z} \psi^{(n)}_{p_z}(x)=\varepsilon_{(n)}(p_z)\psi^{(n)}_{p_z}(x)$, 
where $\hat H_{p_z} =\bm F(\pi_x\to \hat p_x,\pi_y\to -qBx,\pi_z\to p_z)\cdot\bm\sigma$ is a Hamiltonian in the 1-dimensional space of $x$, parametrized by $p_z$, and $\hat p_x=-i\hbar\partial_x$. The non negative integer $n$ is the discrete label for the spectral curves, $\varepsilon_{(n)}(p_z)$, as a function of the continuous parameter $p_z$, which may well be called the generalized Landau levels.
The energy spectrum do not depend on $p_y$, and we have the usual density of states per unit transverse area, $g_\perp={qB\over 2\pi\hbar}$. 
\begin{figure}
\includegraphics[width=9cm]{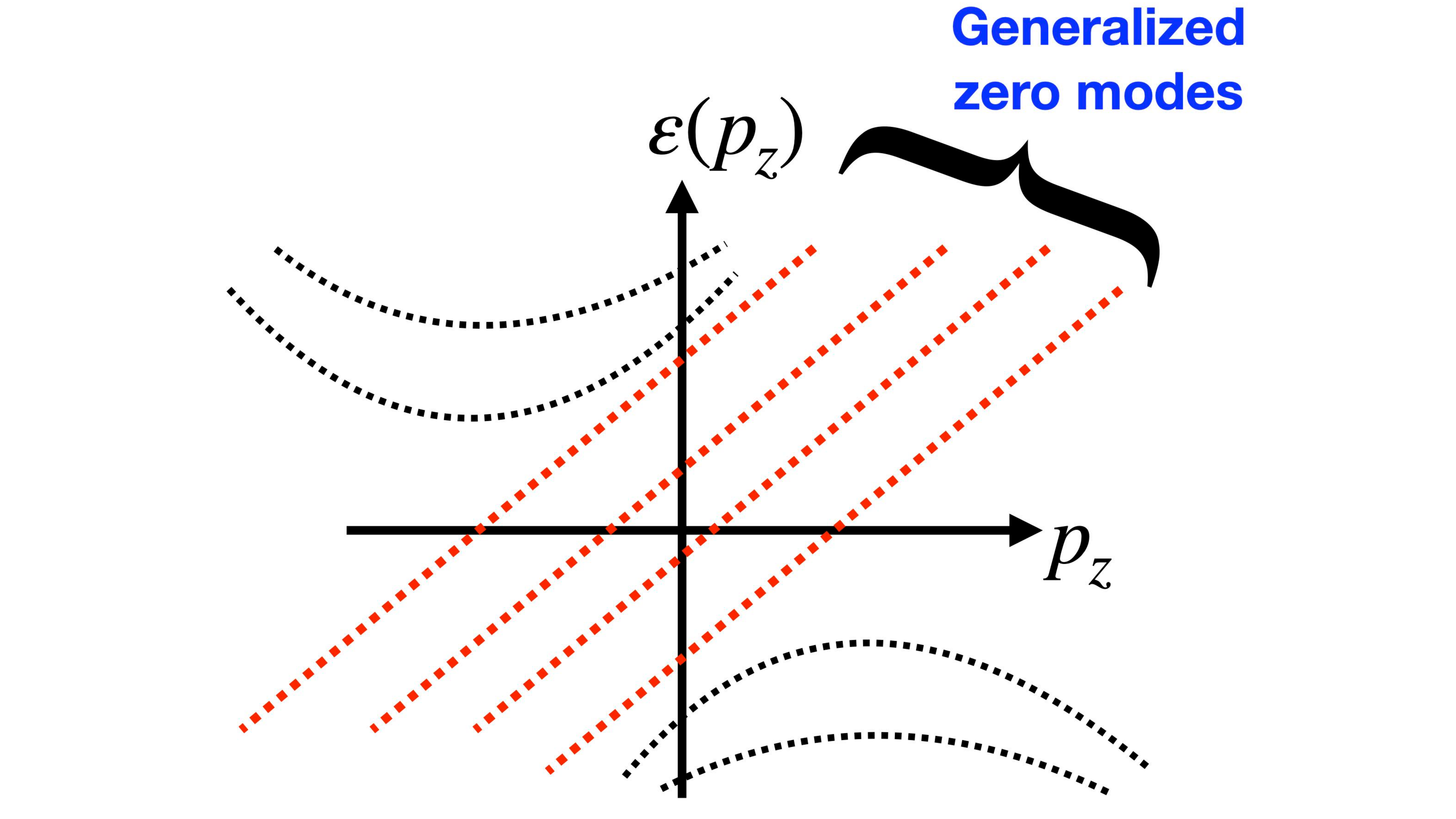}
\caption{A schematic example of spectral curves, $\varepsilon_{(n)}(p_z)$. \label{Fig1}}
\end{figure}
A typical shape of the spectral curves is depicted in FIG. \ref{Fig1}. Using the Feynman-Hellmann theorem, the current expectation value is $\langle \psi^{(n)}_{p_z}|\hat J_z|\psi^{(n)}_{p_z}\rangle=q\langle \psi^{(n)}_{p_z}|{\partial \hat H_{p_z}\over\partial p_z}|\psi^{(n)}_{p_z}\rangle=q{\partial\varepsilon_{(n)}(p_z)\over\partial p_z}$, which implies that for each $n$, the contribution to $\rho_J(\varepsilon)$ is equal to that from a 1-dimensional problem with a Hamiltonian $\hat H_n\equiv\varepsilon_{(n)}(\hat p_z)$, i.e., ${q\over 2\pi\hbar} \sum_\alpha {\varepsilon_{(n)}'(p_\alpha)\over |\varepsilon_{(n)}'(p_\alpha)|}$, that we discussed as a toy example. A non vanishing contribution comes only from the curves with no lower or upper bounds in energy, which we may call the generalized zero modes. We then conclude that $\rho_J(\varepsilon)=g_\perp{q\over 2\pi\hbar}N_0={q^2 B\over (2\pi\hbar)^2}N_0$, where $N_0$ is the signed total number of zero modes of the Hamiltonian $\hat H_{p_z}$, which is a topological property of $\hat H_{p_z}$. Our result for $C_J$ then proves the relation $N_0=N_{\bm F}$.

For a special case of
\be \label{eqn:spham}
\hat H=\bm F_\perp(\bm {\hat\pi_\perp})\cdot\bm\sigma_\perp+\hat p_z \sigma_z,
\ee
 where $(\bm {\hat\pi_\perp},\bm\sigma_\perp)$ have only the transverse components in $(x,y)$ directions, we have $\hat H^2=\hat D_\perp^2+ \hat p_z^2$, where $\hat D_\perp\equiv \bm F_\perp(\bm {\hat\pi_\perp})\cdot\bm\sigma_\perp$ is a generalized Dirac operator in 2-dimensions.
Our expression for $C_J$ then factorizes as \be C_J={1\over\sqrt{\pi}M}{1\over V}{\rm Tr}\left(\hat J_z e^{-\hat H^2/M^2}\right)={q\over 2\pi\hbar}{1\over V_\perp}{\rm Tr}\left(\sigma_z e^{-\hat D_\perp^2/M^2}\right),\ee
where we performed the trace over $z$-dimension, and the last trace is defined only in the transverse 2-dimensions. Noting that $\{\hat D_\perp,\sigma_z\}_+=0$, the trace coincides with the Atiyah-Singer index of the Dirac operator $\hat D_\perp$. Our result then gives a generalized version of the Atiyah-Singer index theorem,
\be
{\rm Index}(\hat D_\perp)={qB\over 2\pi\hbar} N_{\bm F} V_\perp={q \over 2\pi\hbar}N_{\bm F}\int_{R^2_\perp} F_2,
\ee
where $N_{\bm F}\in \pi_1(S^1)={\bf Z}$ is the winding number of the map $\bm p_\perp\to \bm F_\perp(\bm p_\perp)$ at asymptotic infinity, i.e., $S^1_\infty\to S^1_\infty$, which defines the topology of the Dirac operator $\hat D_\perp$.

\section{Application to the chiral magnetic effect}
Our result implies that in the system described by a Hamiltonian $\hat H=\bm F(\bm {\hat\pi})\cdot\bm\sigma$, there is a non vanishing current density along the direction of the magnetic field in grand canonical ensemble,
\be
j_z=C_J\mu={q^2 B\over (2\pi \hbar)^2}N_{\bm F}\mu,
\ee
which is the chiral magnetic effect, that is generalized to arbitrary topological number $N_{\bm F}$.
As shown in Ref.\cite{Yee:2019rot}, the same topological number also appears in the chiral anomaly relation, $\partial_\mu j^\mu={q^3\over (2\pi\hbar)^2} N_{\bm F}(\bm E\cdot\bm B)$, where $j^\mu$ is the charge current density in relativistic notation. This affirms the connection between the chiral magnetic effect and chiral anomaly in the most general case of topology.


In real Weyl semi-metals, as in all condensed matter systems, the global spectrum has a lower bound, and our Theorem I dictates that there is no net persistent current in any thermodynamic ensembles. This is consistent with the Nielsen-Ninomiya theorem that the summation of $N_{\bm F}$ for all Weyl nodes vanishes~\cite{Nielsen}.
Instead, one may consider a quasi equilibrium state, where each Weyl node, labeled by $\alpha$, with a non vanishing $N_{\bm F_\alpha}$, has its own effective chemical potential $\mu_\alpha$, and $j_z={q^2 B\over (2\pi\hbar)^2} \sum_\alpha N_{\bm F_\alpha}\mu_\alpha$ may not vanish.
These quasi equilibrium states would make sense, only if inter-node transitions are slow enough.

\section{A Weyl node with impurities and a magnetic field}
Our general expectation is that $C_J$ is robust under reasonable perturbations, including impurity potentials.
We now study the impact of a perturbing potential on ${{C}_{J}}$ more explicitly.  We considered a number of situations with various perturbations, and proved explicitly that ${{C}_{J}}$ is not affected by these perturbations. We summarize here the main elements and ideas of our proofs (see the Supplemental Materials for full details).

We examined the Hamiltonians with the general form, ${{\hat{H}}_{\lambda }}=\bm{\sigma }\cdot \bm{F}({\bm{\hat{\pi }}})+\lambda \hat{V}$, where $\hat{V}$ is the perturbation and $\lambda$ is a parameter which controls the magnitude of the perturbation.  The perturbations we consider have a general form $\hat{V}(\bm{x})={{V}_{0}}(\bm{x})+\bm{V}(\bm{x})\cdot \bm{\sigma }$ where $\bm{V}$ is a vector valued function representing a spin-dependent potential, and ${{V}_{0}}$ is a spin-independent potential. Recall that in the Landau gauge we have $\bm{\hat{\pi }}=({{{\hat{p}}}_{x}},{{{\hat{p}}}_{y}}-qBx,{{{\hat{p}}}_{z}})$.   The following cases have been examined;
\bear
&& (1)\hspace{.1 cm} {{\hat{H}}_{1\lambda }}={{\bm{\sigma }}_{\bot }}\cdot {{\bm{F}}_{\bot }}( {{{\hat{\bm\pi }}}_{\bot }})+{{\sigma }_{z}}{{\hat{p}}_{z}}+\lambda \hat{V}(x,y)\nonumber\\
&& (2)\hspace{.1 cm}  {{\hat{H}}_{2\lambda }}=\bm{\sigma }\cdot \bm{\hat{\pi }}+\lambda \hat{V}(x,y,z)\nonumber\\
&& (3)\hspace{.1 cm} {{\hat{H}}_{3\lambda }}=\left[ {{\sigma }_{x}}{{ {{{\hat{\pi }}}_{x}} }^{{{n}_{x}}}}+{{\sigma }_{y}}{{ {{{\hat{\pi }}}_{y}} }^{{{n}_{y}}}}+{{\sigma }_{z}}{{ {{{\hat{\pi }}}_{z}} }^{{{n}_{z}}}} \right]+\lambda {{V}_{0}}(x,y,z)\nonumber
\eear
where the $n_i\ge1$ ($i=x,y,z$) in Case 3 are positive integers.
Note that in Case 1, $\hat{V}(x,y)$ is independent of $z$, and the momentum along the $z$ direction, $p_z$, is a good quantum number, which makes the proof particularly simple. This proof is presented in Section 1 of the supplimental materials. 

Our main technique for the proofs is again the $M\to\infty$ limit and the heat kernel expansion.
As an example, we present here an outline of the main analysis for Case 2.	
We first write $\hat{H}_{2\lambda }^{2}={{\bm{\hat{p}}}^{2}}+{{\hat{H}}_{c}}$ where ${{\hat{H}}_{c}}=\hat{H}_{2\lambda }^{2}-{{\bm{\hat{p}}}^{2}}$, and do the heat kernel expansion for $C_J$ treating $\hat{H_c}/M^2$ as a perturbation. Taking the trace in momentum basis, and using the fact that in the large $M$ limit we can ignore position-momentum commutators, the exponential term in the expression for ${{C}_{J}}$ can be Taylor expanded as  ${{{e}}^{-\frac{{{\bm{p}}^{2}}}{{{M}^{2}}}}}{{{e}}^{-\frac{{{{\hat{H}}}_{c}}}{{{M}^{2}}}}}={{{e}}^{-\frac{{{\bm{p}}^{2}}}{{{M}^{2}}}}}\left( 1-\frac{{{{\hat{H}}}_{c}}}{{{M}^{2}}}+\frac{\hat{H}_{c}^{2}}{2!{{M}^{4}}}+\ldots  \right)$. From this we obtain ${{C}_{J}}=\sum_{m=0}^\infty C_{Jm}$ where ${{C}_{Jm}}$ is given by
\be
\frac{{{(-1)}^{m}q}}{{{M}^{( 2m+1)}}m!\sqrt{\pi }}\frac{1}{V}\int{\frac{{{d}^{3}}\bm p}{{{\left( 2\pi \hbar \right)}^{3}}}\int{{{d}^{3}}\bm x}}{{{e}}^{-\frac{{{\bm{p}}^{2}}}{{{M}^{2}}}}}{\rm Tr}_{\rm spin}{\left( {{\sigma }_{z}}\hat{H}_{c}^{m}\right)}.
\ee
Since we are interested in the dependence of ${{C}_{J}}$ on $\hat{V}$, we only consider the terms that include $\hat{V}$. The terms with odd powers of $\bm p$ drop out upon integration, and many terms vanish by spin trace. After this it is found that in the limit $M\to \infty$ the only terms that remain are ${{C}_{J1}}$ and ${{C}_{J2}}$. The $\hat{V}$ dependent parts of these terms are non-zero, however they cancel in the sum ${{C}_{J1}}+{{C}_{J2}}$ quite non-trivially, proving that ${{C}_{J}}$ is indeed independent of $\hat{V}$. Further details are given in Section 2 of the supplimental materials.
Similar steps are used to prove that $C_J$ is independent of $V_0$ for the Case 3 (see Section 3 of the Supplemental Materials for details).

The independence of $C_J$ on the potential is a non-trivial fact, if viewed naively in quantum mechanics. The constant ${{C}_{J}}$ as expressed in Eq.(\ref{eqn:hteqcj}) is  a complicated expression when expanded as a power series of $\hat V$. To obtain the claimed result requires precise cancellation of all terms containing $\hat{V}$.  To demonstrate explicitly that these cancellations do indeed occur, we considered the simple case where ${{\hat{H}}_{1a\lambda }}=\bm{\sigma }\cdot \bm{\hat{\pi }}+\lambda \hat{V}(x,y)$, and showed that the terms in $C_J$ up to second order of $\hat{V}$ do cancel out exactly for any value of $M$. This non-trivial calculation is provided in the Section 4 of the Supplemental Materials.

\section{Implication for the spectral density}
We discuss one immediate consequence of our results on the spectral density of a Weyl node in a magnetic field and impurities.
We first state the following spectral theorem for the Hamiltonians with $\left|{\partial \bm F(\bm {\hat\pi})\over\partial \hat p_z}\right|\le C$ for all normalized states, where $C$ is a non-zero positive constant,
 \be
{\rm Theorem\,\, II:}\quad \rho(\varepsilon)\ge |C_J|/(qC),\,\, {\rm for\,\,all}\,\,\varepsilon.
\ee
An example is
$\hat H_0=\bm F_\perp(\bm {\hat \pi_\perp})\cdot\bm\sigma_\perp+\hat p_z \sigma_z$, where $\left|{\partial \bm F(\bm {\hat \pi})\over\partial \hat p_z}\right|=1$. The proof is based on $|\langle \hat J_z\rangle|=q|\langle {\partial\bm F\over\partial \hat p_z}\cdot\bm \sigma\rangle|\le q \langle|{\partial \bm F(\bm {\hat \pi})\over\partial \hat p_z}|\rangle\le qC$, and we have $|\rho_J(\varepsilon)|={1\over V}\left|{\rm Tr}\left(\hat J_z \delta(\hat H-\varepsilon)\right)\right|\le qC{1\over V}{\rm Tr}\left(\delta(\hat H-\varepsilon)\right)=qC\rho(\varepsilon)$. 
The theorem implies that the energy spectrum has no gap if $\rho_J(\varepsilon)=C_J$ is non vanishing, which is the case for a Weyl node in a magnetic field and impurities.

\section{Discussion}
Our result implies the absence of localization transition for a single Weyl node, as the impurity strength varies, when an external magnetic field is present.
Whether this behavior is smooth in $B\to 0$ limit, or there is a discontinuity at $B=0$ is an interesting question to study.
In real Weyl semi-metals, different Weyl nodes with opposite topological numbers are separated in momentum space, and the inter-node mixings can occur for sufficiently strong short-ranged potentials, which could invalidate our conclusion based on a single Weyl node. 
The robustness of $C_J$ for a single Weyl node shows that the chiral magnetic effect is not affected by any random impurities, as long as inter-node transitions are suppressed. This should be ultimately related to the robustness of chiral anomaly~\cite{Adler:1969er,Lee} that is in effect near each Weyl node. 

The same constant $C_J$ is also responsible for the chiral energy transfer along a magnetic field, $T_{0z}=\int_{-\infty}^{\infty} d\varepsilon (f(\varepsilon)-\Theta(-\varepsilon)){1\over V}{\rm Tr}(\hat v_z \hat H\delta(\hat H-\varepsilon))={1\over q}\int d\varepsilon (f(\varepsilon)-\Theta(-\varepsilon))\varepsilon {1\over V}{\rm Tr}(\hat J_z \delta(\hat H-\varepsilon))={C_J\over q}\int d\varepsilon (f(\varepsilon)-\Theta(-\varepsilon))\varepsilon={C_J\over q}\left({\mu^2\over 2}+{\pi^2 T^2\over 6}\right)={qB N_{\bf F}\over (2\pi\hbar)^2}\left({\mu^2\over 2}+{\pi^2 T^2\over 6}\right)$, where we used $\hat J_z=q\hat v_z$ with the velocity operator $\hat v_z={\partial \hat H\over\partial p_z}$. The chiral energy transfer is equivalent to the chiral vortical effect in time-reversal invariant systems~\cite{Li:2018srq}. 

Our Theorem I is general enough to be applicable to interacting multi-particle systems and quantum field theories of charged particles, as it only relies upon gauge invariance. Moreover, the idea can easily be generalized to produce many similar versions. For example, for any gauge invariant operator $\hat{\cal O}$ which is independent of $\hat\pi_z$ and commutes with $\hat{J_z}$, the weighted spectral density $\rho_{\cal O}(\varepsilon)=\lim_{V\to\infty}{1\over V}{\rm Tr}\left(\hat{\cal O}\hat{J_z}\delta(\hat H-\varepsilon)\right)$
is independent of $\varepsilon$, and hence robust under perturbations. It is an interesting speculation whether this line of thinking might lead to a useful definition of topology for interacting multi-particle systems.


\vskip0.3cm
\section{Acknowledgment}
This work is supported by the U.S. Department of Energy, Office of Science, Office of Nuclear Physics, Grant No. DE-FG0201ER41195.

\bibliographystyle{my-refs}


\begin{widetext}
\vskip 1cm
\begin{center}
\textbf{\large Supplemental Materials}
\end{center}
\vskip 0.5cm

In this supplemental materials, we will work out some of the claims made in the Letter in more detail.  Throughout this discussion we will use $\hbar =1$, the electric charge $q=1$, and $\bm{\hat{\pi }}=\left( {{{\hat{p}}}_{x}},{{{\hat{p}}}_{y}}-Bx,{{{\hat{p}}}_{z}} \right)$.

\begin{center}
{\bf Section 1}  
\end{center}

Consider the Hamiltonian ${{\hat{H}}_{1\lambda }}$ which was defined in Case 1 of the Letter and is given here for convenience,
\be
{{\hat{H}}_{1\lambda }}={{\bm{\sigma }}_{\bot }}\cdot {{\bm{F}}_{\bot }}\left( {{{\hat{\pi }}}_{\bot }} \right)+{{\sigma }_{z}}{{\hat{p}}_{z}}+\lambda \hat{V}\left( x,y \right),
\ee	
where $\hat{V}\left( x,y \right)={{V}_{0}}\left( x,y \right)+\bm{\sigma }\cdot \bm{V}\left( x,y \right)$ and we assume that $\hat{V}\left( x,y \right)\to 0$ at the boundaries.  Note that both  ${{\bm{\sigma }}_{\bot }}\cdot {{\bm{F}}_{\bot }}\left( {{{\hat{\pi }}}_{\bot }} \right)$and $\hat{V}\left( x,y \right)$ are independent of $z$. We want to prove that ${{C}_{J}}$ is not dependent on $\hat{V}$ or, more specifically, not dependent on $\lambda$. 
	
From Eq. (8) of the Letter, ${{C}_{J}}$ is given by
\be 
{{C}_{J}}=\frac{1}{M\sqrt{\pi }}\frac{1}{V}Tr\left\{ {{\sigma }_{z}}{{e}^{-\frac{\hat{H}_{1\lambda }^{2}}{{{M}^{2}}}}} \right\}.
\ee	
This is equivalent to
\be 
{{C}_{J}}=\frac{1}{2\pi V}Tr\left\{ \int\limits_{-\infty }^{+\infty }{{{\sigma }_{z}}{{e}^{i{{{\hat{H}}}_{\lambda }}t}}{{e}^{-\frac{1}{4}{{M}^{2}}{{t}^{2}}}}dt} \right\},
\ee	
where we have used the relationship $\int\limits_{-\infty }^{+\infty }{{{e}^{-\left( {{{M}^{2}}}/{4}\; \right){{t}^{2}}}}{{e}^{ibt}}dt}=\sqrt{\left( {4\pi }/{{{M}^{2}}}\; \right)}{{e}^{-\left( {{{b}^{2}}}/{{{M}^{2}}}\; \right)}}$.
Since ${{\bm{F}}_{\bot }}\left( {{{\hat{\pi }}}_{\bot }} \right)$ and $\lambda \hat{V}$ are independent of $z$ the effect of taking the trace over the $z$-dimension is simply to replace the operator ${{\hat{p}}_{z}}$ with the scalar ${{p}_{z}}$.  Doing this yields
\be 
{{C}_{J}}=\frac{1}{2\pi V}Tr\left\{ \int\limits_{-\infty }^{+\infty }{dt{{e}^{-\frac{1}{4}{{M}^{2}}{{t}^{2}}}}\int\limits_{-W}^{+W}{\frac{{{L}_{z}}d{{p}_{z}}}{2\pi }}{{\sigma }_{z}}{{e}^{i\left( {{\bm{F}}_{\bot }}\left( {{{\hat{\pi }}}_{\bot }} \right)\cdot {{\bm{\sigma }}_{\bot }}+{{\sigma }_{z}}\cdot {{p}_{z}}+\lambda \hat{V} \right)t}}} \right\},
\ee	
where $W\to \infty$ limit will be taken, and the trace is now over the $x$,$y$-dimensions and the spinors.  From the above we obtain
\be 
{{C}_{J}}=\frac{{{L}_{z}}}{{{\left( 2\pi  \right)}^{2}}V}Tr\left\{ \int\limits_{-\infty }^{+\infty }{dt{{e}^{-\frac{1}{4}{{M}^{2}}{{t}^{2}}}}\int\limits_{-W}^{+W}{d{{p}_{z}}}\frac{\partial }{it\partial {{p}_{z}}}{{e}^{i\left( {{\bm{F}}_{\bot }}\left( {{{\hat{\pi }}}_{\bot }} \right)\cdot {{\bm{\sigma }}_{\bot }}+{{\sigma }_{z}}\cdot {{p}_{z}}+\lambda \hat{V} \right)t}}} \right\}.
\ee	
Next perform the integration over ${{p}_{z}}$ to obtain
\be 
{{C}_{J}}=\frac{{{L}_{z}}}{{{\left( 2\pi  \right)}^{2}}V}Tr\left\{ \int\limits_{-\infty }^{+\infty }{\frac{dt}{it}{{e}^{-\frac{1}{4}{{M}^{2}}{{t}^{2}}}}\left[ {{e}^{i\left( {{\bm{F}}_{\bot }}\left( {{{\hat{\pi }}}_{\bot }} \right)\cdot {{\bm{\sigma }}_{\bot }}+{{\sigma }_{z}}\cdot W+\lambda \hat{V} \right)t}}-{{e}^{i\left( {{\bm{F}}_{\bot }}\left( {{{\hat{\pi }}}_{\bot }} \right)\cdot {{\bm{\sigma }}_{\bot }}-{{\sigma }_{z}}\cdot W+\lambda \hat{V} \right)t}} \right]} \right\}.
\ee
We next take the derivative of both sides with respect to $\lambda$ to obtain
\be 
\frac{\partial {{C}_{J}}}{\partial \lambda }=\frac{{{L}_{z}}}{{{\left( 2\pi  \right)}^{2}}V}Tr\left\{ \hat{V}\int\limits_{-\infty }^{+\infty }{dt{{e}^{-\frac{1}{4}{{M}^{2}}{{t}^{2}}}}\left[ {{e}^{i\left( {{\bm{F}}_{\bot }}\left( {{{\hat{\pi }}}_{\bot }} \right)\cdot {{\bm{\sigma }}_{\bot }}+{{\sigma }_{z}}\cdot W+\lambda \hat{V} \right)t}}-{{e}^{i\left( {{\bm{F}}_{\bot }}\left( {{{\hat{\pi }}}_{\bot }} \right)\cdot {{\bm{\sigma }}_{\bot }}-{{\sigma }_{z}}\cdot W+\lambda \hat{V} \right)t}} \right]} \right\},
\ee	
and performing the integration, we obtain
\be \label{eqn:eqn08}
\frac{\partial {{C}_{J}}}{\partial \lambda }=\frac{{{L}_{z}}2\sqrt{\pi }}{{{\left( 2\pi  \right)}^{2}}MV}Tr\left\{ \hat{V}\left[ {{e}^{-\frac{{{{\hat{H}}}_{1\lambda }}{{\left( W \right)}^{2}}}{{{M}^{2}}}}}-{{e}^{-\frac{{{{\hat{H}}}_{1\lambda }}{{\left( -W \right)}^{2}}}{{{M}^{2}}}}} \right] \right\},
\ee	
where ${{\hat{H}}_{1\lambda }}{{\left( \pm W \right)}^{2}}={{\left( {{\bm{F}}_{\bot }}\left( {{{\hat{\pi }}}_{\bot }} \right)\cdot {{\bm{\sigma }}_{\bot }}\pm {{\sigma }_{z}}W+\lambda \hat{V} \right)}^{2}}={{\left( {{\bm{F}}_{\bot }}\left( {{{\hat{\pi }}}_{\bot }} \right)\cdot {{\bm{\sigma }}_{\bot }}+\lambda \hat{V} \right)}^{2}}\pm 2W\lambda \left( {{\sigma }_{z}}{{V}_{0}}+{{V}_{z}} \right)+{{W}^{2}}$.  Then in the limit $W\to \infty$ the exponentials in Eq. (\ref{eqn:eqn08}) approach zero, and therefore the quantity $\partial {{C}_{J}}/\partial \lambda \ \to 0$, which proves that ${{C}_{J}}$ is independent of $\hat{V}\left( x,y \right)$.

\begin{center}
{\bf Section 2}  
\end{center}

We next consider a case where the perturbation $\hat{V}$ has $z$-dependence, so that the conditions of the proof given in Section 1 are not met.  In this case we will use the fact that ${{C}_{J}}$ is independent of $M$ and operate in the large $M$ limit where $M\to \infty $. The effect of this will be demonstrated in the following example.  Consider the Hamiltonian ${{\hat{H}}_{2\lambda }}$ which is specified in Case 2 of the Letter,
\be 
{{\hat{H}}_{2\lambda }}={{\hat{H}}_{0}}+\lambda \hat{V}\left( x,y,z \right),\quad {{\hat{H}}_{0}}=\bm{\sigma }\cdot \bm{\hat{\pi }},
\ee	
where $\hat{V}\left( x,y,z \right)={{V}_{0}}\left( x,y,z \right)+\bm{\sigma }\cdot \bm{V}\left( x,y,z \right)$.  From this we obtain $\hat{H}_{2\lambda }^{2}={{\bm{\hat{p}}}^{2}}+{{\hat{H}}_{c}}$ where,	
\be 
{{\hat{H}}_{c}}=\left[ \left( {{{\hat{H}}}_{0}}\hat{V}+\hat{V}{{{\hat{H}}}_{0}} \right)+{{{\hat{V}}}^{2}} \right]+\left( {{B}^{2}}{{x}^{2}}-2Bx{{{\hat{p}}}_{y}} \right)-B{{\sigma }_{z}},
\ee	
and we have set $\lambda =1$.  We want to evaluate  
\be 
{{C}_{J}}=\frac{1}{MV\sqrt{\pi }}Tr\left\{ {{\sigma }_{z}}{{e}^{-\frac{\hat{H}_{2\lambda }^{2}}{{{M}^{2}}}}} \right\}.
\ee	
First take the trace in the momentum basis wave functions $\left| \psi \left( \bm{p} \right) \right\rangle =\frac{1}{\sqrt{V}}{{e}^{i\bm{p}\cdot \bm{x}}}$, and note that, as will be discussed below, in the large $M$ limit ($M\to \infty $) we can ignore commutators between the momentum operators and $\hat{V}$, and the momentum operator can be replaced by its eigenvalues.  The result of all this is 
\be 
{{C}_{J}}=\frac{1}{M\sqrt{\pi }}\frac{1}{V}T{{r}_{spin}}\left\{ {{\sigma }_{z}}\int{\frac{{{d}^{3}}p}{{{\left( 2\pi  \right)}^{3}}}\int{{{d}^{3}}x}}{{\operatorname{e}}^{-\frac{\left( {{\bm{p}}^{2}}+{{{\hat{H}}}_{c}} \right)}{{{M}^{2}}}}} \right\}.
\ee	
We can write the exponential in the above in the Taylor series as   ${{\operatorname{e}}^{-\frac{{{\bm{p}}^{2}}}{{{M}^{2}}}}}{{\operatorname{e}}^{-\frac{{{{\hat{H}}}_{c}}}{{{M}^{2}}}}}={{\operatorname{e}}^{-\frac{{{\bm{p}}^{2}}}{{{M}^{2}}}}}\left( 1-\frac{{{{\hat{H}}}_{c}}}{{{M}^{2}}}+\frac{\hat{H}_{c}^{2}}{2!{{M}^{4}}}+\ldots  \right)$.  When this is substituted back into the above equation we obtain ${{C}_{J}}={{C}_{J0}}+{{C}_{J1}}+{{C}_{J2}}+\ldots $, where		
\be 
{{C}_{Jm}}=\frac{{{\left( -1 \right)}^{m}}}{{{M}^{\left( 2m+1 \right)}}\sqrt{\pi }}\frac{1}{V}T{{r}_{spin}}\left\{ {{\sigma }_{z}}\int{\frac{{{d}^{3}}p}{{{\left( 2\pi  \right)}^{3}}}\int{{{d}^{3}}x}}{{\operatorname{e}}^{-\frac{{{\bm{p}}^{2}}}{{{M}^{2}}}}}\frac{\hat{H}_{c}^{m}}{m!} \right\}.
\ee
The reason we want to work in the large $M$ limit is that it results in great simplification in evaluating the above expressions.  To show this explicitly, consider the term  ${{C}_{J4}}$.  First note that all terms with an odd order of ${{p}_{i}}$ drop out.  Then note that the highest order of ${{p}_{i}}$ in $C_{J4}$ is ${{p}_{i}}^{4}$.  Therefore the maximum value the integration over the momentum can have is $\int{{{d}^{3}}p{{\operatorname{e}}^{-\frac{{{\bm{p}}^{2}}}{{{M}^{2}}}}}p_{z}^{4}}\sim{{M}^{7}}$.  Therefore ${{C}_{J4}}\sim{{M}^{7}}/{{M}^{9}}\ \to 0$ as $M\to \infty $.  It can be shown that all ${{C}_{Jm}}$ with $m\ge 3$ drop out in the large $M$ limit by similar power counting.  This is also why we can ignore commutators between the momentum operators and $\hat{V}$.  The action of taking the commutator drops the order of the momentum by one and in the large $M$ limit these terms will drop out.

Since we are interested in the dependence of ${{C}_{J}}$ on $\hat{V}$ we will only consider the parts of these terms that include $\hat{V}$.   Therefore we only need to consider the terms ${{C}_{J1}}$ and ${{C}_{J2}}$.  First evaluate ${{C}_{J1}}$ which is given as
\be 
{{C}_{J1}}=\frac{-1}{{{M}^{3}}\sqrt{\pi }V}T{{r}_{spin}}\left\{ {{\sigma }_{z}}\int{\frac{{{d}^{3}}p}{{{\left( 2\pi  \right)}^{3}}}\int{{{d}^{3}}x}}{{\operatorname{e}}^{-\frac{{{\bm{p}}^{2}}}{{{M}^{2}}}}}\left[ \left[ \left( {{{\hat{H}}}_{0}}\hat{V}+\hat{V}{{{\hat{H}}}_{0}} \right)+{{{\hat{V}}}^{2}} \right]+\left( {{B}^{2}}{{x}^{2}}-2Bx{{p}_{y}} \right)-B{{\sigma }_{z}} \right] \right\}.
\ee		
Next remove all terms that do not include $\hat{V}$ to obtain
\be 
{{C}_{J1}}\left( {\hat{V}} \right)=\frac{-1}{{{M}^{3}}\sqrt{\pi }V}T{{r}_{spin}}\left\{ {{\sigma }_{z}}\int{\frac{{{d}^{3}}p}{{{\left( 2\pi  \right)}^{3}}}\int{{{d}^{3}}x}}{{\operatorname{e}}^{-\frac{{{\bm{p}}^{2}}}{{{M}^{2}}}}}\left[ \left( {{{\hat{H}}}_{0}}\hat{V}+\hat{V}{{{\hat{H}}}_{0}} \right)+{{{\hat{V}}}^{2}} \right] \right\}.
\ee	
We note that all terms that are to the odd power in ${{p}_{i}}$ integrate out to zero, and we obtain
\be 
{{C}_{J1}}\left( {\hat{V}} \right)=\frac{-1}{{{M}^{3}}\sqrt{\pi }V}T{{r}_{spin}}\left\{ {{\sigma }_{z}}\int{\frac{{{d}^{3}}p}{{{\left( 2\pi  \right)}^{3}}}\int{{{d}^{3}}x}}{{\operatorname{e}}^{-\frac{{{\bm{p}}^{2}}}{{{M}^{2}}}}}{{{\hat{V}}}^{2}} \right\}.
\ee	
Next use ${{\hat{V}}^{2}}=V_{0}^{2}+2{{V}_{0}}\bm{\sigma }\cdot \bm{V}+{{\left| \bm{V} \right|}^{2}}$ and take the trace over spinors and integrate w.r.t to the ${{p}_{i}}$ to obtain,
\be \label{eqn:eq56}
{{C}_{J1}}\left( {\hat{V}} \right)=\frac{-1}{{{M}^{3}}\sqrt{\pi }V}\left\{ \int{\frac{{{d}^{3}}p}{{{\left( 2\pi  \right)}^{3}}}\int{{{d}^{3}}x}}{{\operatorname{e}}^{-\frac{{{\bm{p}}^{2}}}{{{M}^{2}}}}}4{{V}_{0}}{{V}_{z}} \right\}=\frac{-1}{2{{\pi }^{2}}V}\left\{ \int{{{d}^{3}}x}\left( {{V}_{0}}{{V}_{z}} \right) \right\}.
\ee	
We next examine ${{C}_{J2}}$,	
\be 
{{C}_{J2}}=\frac{1}{2{{M}^{5}}\sqrt{\pi }V}T{{r}_{spin}}\left\{ {{\sigma }_{z}}\int{\frac{{{d}^{3}}p}{{{\left( 2\pi  \right)}^{3}}}\int{{{d}^{3}}x}}{{\operatorname{e}}^{-\frac{{{\bm{p}}^{2}}}{{{M}^{2}}}}}{{\left[ \left( {{{\hat{H}}}_{0}}\hat{V}+\hat{V}{{{\hat{H}}}_{0}} \right)+{{{\hat{V}}}^{2}}+\left( {{B}^{2}}{{x}^{2}}-2Bx{{p}_{y}} \right)-B{{\sigma }_{z}} \right]}^{2}} \right\}.
\ee													
We neglect all terms with no dependence on $\hat{V}$, and also some terms will be eliminated when we take the trace over spinors.  In addition all terms with an odd order of ${{p}_{i}}$ will integrate out to zero.  Note that all surviving terms must have at least second power of ${{p}_{i}}$, otherwise they will go to zero when the $M\to \infty $ limit is taken.  This leaves us with	
\be 
{{C}_{J2}}\left( {\hat{V}} \right)=\frac{1}{2{{M}^{5}}\sqrt{\pi }V}T{{r}_{spin}}\left\{ {{\sigma }_{z}}\int{\frac{{{d}^{3}}p}{{{\left( 2\pi  \right)}^{3}}}\int{{{d}^{3}}x}}{{\operatorname{e}}^{-\frac{{{\bm{p}}^{2}}}{{{M}^{2}}}}}{{\left( {{{\hat{H}}}_{0}}\hat{V}+\hat{V}{{{\hat{H}}}_{0}} \right)}^{2}} \right\},
\ee	
which is in more detail
\be 
{{C}_{J2}}\left( {\hat{V}} \right)=\frac{1}{2{{M}^{5}}\sqrt{\pi }V}T{{r}_{spin}}\left\{ {{\sigma }_{z}}\int{\frac{{{d}^{3}}p}{{{\left( 2\pi  \right)}^{3}}}\int{{{d}^{3}}x}}{{\operatorname{e}}^{-\frac{{{\bm{p}}^{2}}}{{{M}^{2}}}}}\left[ \begin{aligned}
  & \left( {{{\hat{H}}}_{0a}}+{{\sigma }_{z}}{{p}_{z}} \right)\hat{V}\left( {{{\hat{H}}}_{0a}}+{{\sigma }_{z}}{{p}_{z}} \right)\hat{V} \\ 
 & +\left( {{{\hat{H}}}_{0a}}+{{\sigma }_{z}}{{p}_{z}} \right)\hat{V}\hat{V}\left( {{{\hat{H}}}_{0a}}+{{\sigma }_{z}}{{p}_{z}} \right) \\ 
 & +\hat{V}\left( {{{\hat{H}}}_{0a}}+{{\sigma }_{z}}{{p}_{z}} \right)\hat{V}\left( {{{\hat{H}}}_{0a}}+{{\sigma }_{z}}{{p}_{z}} \right) \\ 
 & +\hat{V}\left( {{{\hat{H}}}_{0a}}+{{\sigma }_{z}}{{p}_{z}} \right)\left( {{{\hat{H}}}_{0a}}+{{\sigma }_{z}}{{p}_{z}} \right)\hat{V} \\ 
\end{aligned} \right] \right\},
\ee
where ${{\hat{H}}_{0a}}={{\sigma }_{x}}{{\hat{p}}_{x}}+{{\sigma }_{y}}\left( {{{\hat{p}}}_{y}}-Bx \right)$.  After some manipulation, we have
\be 
{{C}_{J2}}\left( {\hat{V}} \right)=\frac{1}{2{{M}^{5}}\sqrt{\pi }V}T{{r}_{spin}}\left\{ {{\sigma }_{z}}\int{\frac{{{d}^{3}}p}{{{\left( 2\pi  \right)}^{3}}}\int{{{d}^{3}}x}}{{\operatorname{e}}^{-\frac{{{\bm{p}}^{2}}}{{{M}^{2}}}}}\left( \begin{aligned}
  & \left[ \begin{aligned}
  & {{{\hat{H}}}_{0a}}\hat{V}{{{\hat{H}}}_{0a}}\hat{V}+{{{\hat{H}}}_{0a}}\hat{V}\hat{V}{{{\hat{H}}}_{0a}} \\ 
 & +\hat{V}{{{\hat{H}}}_{0a}}\hat{V}{{{\hat{H}}}_{0a}}+\hat{V}{{{\hat{H}}}_{0a}}{{{\hat{H}}}_{0a}}\hat{V} \\ 
\end{aligned} \right] \\ 
 & +p_{z}^{2}\left[ \begin{aligned}
  & {{\sigma }_{z}}\hat{V}{{\sigma }_{z}}\hat{V}+{{\sigma }_{z}}\hat{V}\hat{V}{{\sigma }_{z}} \\ 
 & +\hat{V}{{\sigma }_{z}}\hat{V}{{\sigma }_{z}}+\hat{V}{{\sigma }_{z}}{{\sigma }_{z}}\hat{V} \\ 
\end{aligned} \right] \\ 
\end{aligned} \right) \right\}.
\ee
Next use the commutative property of trace and the fact that ${{\sigma }_{z}}{{\hat{H}}_{0a}}=-{{\hat{H}}_{0a}}{{\sigma }_{z}}$ to eliminate the term $\left[ {{{\hat{H}}}_{0a}}\hat{V}{{{\hat{H}}}_{0a}}\hat{V}+\hat{V}{{{\hat{H}}}_{0a}}\hat{V}{{{\hat{H}}}_{0a}} \right]$.
Also we can replace ${{\hat{H}}_{0a}}\hat{V}\hat{V}{{\hat{H}}_{0a}}$ with $-{{\hat{H}}_{0a}}{{\hat{H}}_{0a}}\hat{V}\hat{V}$.  From all this we obtain
\be 
{{C}_{J2}}\left( {\hat{V}} \right)=\frac{1}{2{{M}^{5}}\sqrt{\pi }V}T{{r}_{spin}}\left\{ \int{\frac{{{d}^{3}}p}{{{\left( 2\pi  \right)}^{3}}}\int{{{d}^{3}}x}}{{\operatorname{e}}^{-\frac{{{\bm{p}}^{2}}}{{{M}^{2}}}}}\left( \begin{aligned}
  & {{\sigma }_{z}}\left[ \begin{aligned}
  & -{{{\hat{H}}}_{0a}}{{{\hat{H}}}_{0a}}\hat{V}\hat{V} \\ 
 & +\hat{V}{{{\hat{H}}}_{0a}}{{{\hat{H}}}_{0a}}\hat{V} \\ 
\end{aligned} \right] \\ 
 & +4{{\sigma }_{z}}p_{z}^{2}\hat{V}\hat{V} \\ 
\end{aligned} \right) \right\}.
\ee	
Since only $p_{i}^{2}$ terms survive in the large $M$ limit, we can replace ${{\hat{H}}_{0a}}{{\hat{H}}_{0a}}\to p_{x}^{2}+p_{y}^{2}$.  The result is that the ${{\hat{H}}_{0a}}{{\hat{H}}_{0a}}$ terms in the above expression cancel out.  Therefore we obtain
\be 
{{C}_{J2}}\left( {\hat{V}} \right)=\frac{1}{2{{M}^{5}}\sqrt{\pi }V}T{{r}_{spin}}\left\{ \int{\frac{{{d}^{3}}p}{{{\left( 2\pi  \right)}^{3}}}\int{{{d}^{3}}x}}{{\operatorname{e}}^{-\frac{{{\bm{p}}^{2}}}{{{M}^{2}}}}}\left( 4{{\sigma }_{z}}p_{z}^{2}{{{\hat{V}}}^{2}} \right) \right\}.
\ee
Using ${{\hat{V}}^{2}}=V_{0}^{2}+2{{V}_{0}}\bm{\sigma }\cdot \bm{V}+{{\left| \bm{V} \right|}^{2}}$ and taking the trace, we have
\be 
{{C}_{J2}}\left( {\hat{V}} \right)=\frac{1}{2{{M}^{5}}\sqrt{\pi }V}\left\{ \int{\frac{{{d}^{3}}p}{{{\left( 2\pi  \right)}^{3}}}\int{{{d}^{3}}x}}{{\operatorname{e}}^{-\frac{{{\bm{p}}^{2}}}{{{M}^{2}}}}}\left( 16p_{z}^{2}{{V}_{0}}{{V}_{z}} \right) \right\},
\ee
and integrating over the ${{p}_{i}}$ gives
\be 
{{C}_{j2}}\left( {\hat{V}} \right)=\frac{1}{2{{\pi }^{2}}V}\int{{{d}^{3}}x}\left( {{V}_{0}}{{V}_{z}} \right).
\ee	
Use this result along with Eq. (\ref{eqn:eq56}) to show that ${{C}_{J1}}\left( {\hat{V}} \right)+{{C}_{J2}}\left( {\hat{V}} \right)$ equals zero.  Therefore we have proved that the part of ${{C}_{J}}$ that is dependent on $\hat{V}$ equals zero.

\begin{center}
{\bf Section 3}  
\end{center}

We will consider the Hamiltonian ${{\hat{H}}_{3\lambda }}$ which is defined in Case 3 in the Letter and is written below,
\be 
{{\hat{H}}_{3\lambda }}={{\hat{H}}_{0}}+\lambda {{V}_{0}}\left( x,y,z \right),\quad {{\hat{H}}_{0}}=\left[ {{\sigma }_{x}}{{{\hat{p}}}_{x}}^{{{n}_{x}}}+{{\sigma }_{y}}{{\left( {{{\hat{p}}}_{y}}-Bx \right)}^{{{n}_{y}}}}+{{\sigma }_{z}}{{{\hat{p}}}_{z}}^{{{n}_{z}}} \right],
\ee
where the ${{n}_{x}}$, ${{n}_{y}}$, ${{n}_{z}}$ are positive integers.
We want to show that ${{C}_{J}}$ is independent of ${{V}_{0}}\left( x,y,z \right)$.  We will proceed as in the previous section and operate in the large $M$ limit.   In the following discussion we set $\lambda =1$.  Refer to Eq. (8) in the Letter and note that ${{\hat{J}}_{z}}=\partial {{\hat{H}}_{3\lambda }}/\partial {{\hat{p}}_{z}}=\ {{\sigma }_{z}}{{\hat{p}}_{z}}^{{{n}_{z}}-1}$ to obtain
\be \label{eqn:eqn027}
{{C}_{J}}=\frac{1}{MV\sqrt{\pi }}Tr\left\{ {{\sigma }_{z}}{{{\hat{p}}}_{z}}^{{{n}_{z}}-1}\exp \left( -\frac{{{{\hat{H}}}_{3\lambda }}^{2}}{{{M}^{2}}} \right) \right\}.
\ee
We have the following expression,
\be \label{eqn:eqn028}
\hat{H}_{3\lambda }^{2}={{\left( {{{\hat{H}}}_{0}}+{{V}_{0}} \right)}^{2}}={{\hat{H}}_{0}}^{2}+\left[ {{{\hat{H}}}_{0}}{{V}_{0}}+{{V}_{0}}{{{\hat{H}}}_{0}} \right]+{{V}_{0}}^{2},
\ee	
and we next write
\be 
{{\hat{H}}_{0}}^{2}={{\hat{H}}_{a0}}+i{{\sigma }_{z}}{{\hat{H}}_{a1}},
\ee	
where $ 
{{\hat{H}}_{a0}}={{\hat{p}}_{x}}^{2{{n}_{x}}}+{{\left( {{{\hat{p}}}_{y}}-Bx \right)}^{2{{n}_{y}}}}+{{\hat{p}}_{z}}^{2{{n}_{z}}}$
and $
{{\hat{H}}_{a1}}=\left[ {{{\hat{p}}}_{x}}^{{{n}_{x}}},{{\left( {{{\hat{p}}}_{y}}-Bx \right)}^{{{n}_{y}}}} \right]\sim iB{{\hat{p}}_{x}}^{{{n}_{x}}-1}{{\hat{p}}_{y}}^{{{n}_{y}}-1}$, 
where the lower order terms in momentum are discarded because they don't contribute to the following calculations in the large $M$ limit.
Note also that ${{\left( {{{\hat{p}}}_{y}}-Bx \right)}^{2{{n}_{y}}}}={{\left( {{{\hat{p}}}_{y}} \right)}^{2{{n}_{y}}}}+{{\hat{H}}_{a2}}$, 
where $
{{\hat{H}}_{a2}}=\left[ {{a}_{1}}{{{\hat{p}}}_{y}}^{2{{n}_{y}}-1}Bx+{{a}_{2}}{{{\hat{p}}}_{y}}^{2{{n}_{y}}-2}{{\left( Bx \right)}^{2}}+\ldots  \right]$ 
with constants ${{a}_{i}}$ that can be derived using the binomial theorem.
Use all this to obtain
\be 
{{\hat{H}}_{0}}^{2}={{\hat{H}}_{00}}+{{\hat{H}}_{a2}}+i{{\sigma }_{z}}{{\hat{H}}_{a1}},\ee
where ${{\hat{H}}_{00}}={{\hat{p}}_{x}}^{2{{n}_{x}}}+{{\hat{p}}_{y}}^{2{{n}_{y}}}+{{\hat{p}}_{z}}^{2{{n}_{z}}}$.
Use this in Eq. (\ref{eqn:eqn028}) to obtain $
\hat{H}_{3\lambda }^{2}={{\hat{H}}_{00}}+{{\hat{H}}_{b}}$, 
where ${{\hat{H}}_{b}}=\left[ {{{\hat{H}}}_{a2}}+i{{\sigma }_{z}}{{{\hat{H}}}_{a1}}+\left( {{{\hat{H}}}_{0}}{{V}_{0}}+{{V}_{0}}{{{\hat{H}}}_{0}} \right)+{{V}_{0}}^{2} \right]$.
Then Eq. (\ref{eqn:eqn027}) becomes \be
{{C}_{J}}=\frac{1}{MV\sqrt{\pi }}Tr\left\{ {{\sigma }_{z}}{{{\hat{p}}}_{z}}^{{{n}_{z}}-1}\exp \left( -\frac{{{{\hat{H}}}_{00}}+{{{\hat{H}}}_{b}}}{{{M}^{2}}} \right) \right\}.\ee
Since we are operating in the large $M$ limit ($M\to \infty$) we can ignore commutators between the momentum operators and ${{V}_{0}}$, and we can do the expansion,
\be 
\exp \left( -\frac{{{{\hat{H}}}_{00}}+{{{\hat{H}}}_{b}}}{{{M}^{2}}} \right)\sim\exp \left( -\frac{{{{\hat{H}}}_{00}}}{{{M}^{2}}} \right)\left\{ 1-\frac{{{{\hat{H}}}_{b}}}{{{M}^{2}}}+\frac{1}{2!}\frac{{{{\hat{H}}}_{b}}^{2}}{{{M}^{4}}}+\ldots  \right\}.
\ee
We will proceed as in the last section to evaluate the trace using the momentum basis wave functions $\left| {{\psi }_{\bm{p}}} \right\rangle =\frac{1}{\sqrt{V}}{{e}^{i\bm{p}\cdot \bm{x}}}$.  In this case we obtain ${{C}_{J}}={{C}_{J0}}+{{C}_{J1}}+{{C}_{J2}}+\ldots$ where,
\be 
{{C}_{Jm}}=\frac{{{\left( -1 \right)}^{m}}}{{{M}^{\left( 2m+1 \right)}}\sqrt{\pi }}\frac{1}{V}T{{r}_{spin}}\left\{ {{\sigma }_{z}}{{p}_{z}}^{{{n}_{z}}-1}\int{\frac{{{d}^{3}}p}{{{\left( 2\pi  \right)}^{3}}}{{\operatorname{e}}^{-\frac{{{p}_{x}}^{2{{n}_{x}}}+{{p}_{y}}^{2{{n}_{y}}}+{{p}_{z}}^{2{{n}_{z}}}}{{{M}^{2}}}}}\int{{{d}^{3}}x}}\frac{\hat{H}_{b}^{m}}{m!} \right\}.
\ee
In the limit $M\to \infty$ we can use the following approximation $\int\limits_{-\infty }^{+\infty }{dp{{e}^{-\frac{{{p}^{2n}}}{{{M}^{2}}}}}{{p}^{a}}}\sim\int\limits_{-{{M}^{\left( 1/n\  \right)}}}^{+{{M}^{\left( 1/n\  \right)}}}{{{p}^{a}}dp}$ for a positive integer $n$, which gives the right power counting in $M$.  Therefore we obtain
\be 
{{C}_{Jm}}\sim \frac{{(-1)}^{m}}{V{{M}^{2m+1}}}T{{r}_{spin}}\int{D\bm{p}}\int{{{d}^{3}}x}\left\{ {{\sigma }_{z}}{{p}_{z}}^{{{n}_{z}}-1}\int{{{d}^{3}}x\frac{\hat{H}_{b}^{m}}{m!}} \right\},
\ee	
where we define$\int{D\bm{p}}=\int\limits_{-{{M}^{\left( {1}/{{{n}_{x}}}\; \right)}}}^{+{{M}^{\left( {1}/{{{n}_{x}}}\; \right)}}}{d{{p}_{x}}}\int\limits_{-{{M}^{\left( {1}/{{{n}_{y}}}\; \right)}}}^{+{{M}^{\left( {1}/{{{n}_{y}}}\; \right)}}}{d{{p}_{y}}}\int\limits_{-{{M}^{\left( {1}/{{{n}_{z}}}\; \right)}}}^{+{{M}^{\left( {1}/{{{n}_{z}}}\; \right)}}}{d{{p}_{z}}}$.  For non-negative even integers $a$, $b$, and $c$, we have 
\be \label{eqn:eqn040}
\int{D\bm{p}}\left( p_{x}^{a}p_{y}^{b}p_{z}^{c} \right)\sim {{M}^{\left[ {\left( a+1 \right)}/{{{n}_{x}}}\; \right]}}{{M}^{\left[ {\left( b+1 \right)}/{{{n}_{y}}}\; \right]}}{{M}^{\left[ {\left( c+1 \right)}/{{{n}_{z}}}\; \right]}},
\ee	
while if any of $a$, $b$, or $c$ are odd then the above quantity will be zero.
Let us now evaluate ${{C}_{J1}}$,
\be 
{{C}_{J1}}\sim-\frac{1}{V{{M}^{3}}}T{{r}_{spin}}\int{D\bm{p}\int{{{d}^{3}}x}}\left\{ {{\sigma }_{z}}{{p}_{z}}^{{{n}_{z}}-1}\left[ {{{\hat{H}}}_{a2}}+i{{\sigma }_{z}}{{{\hat{H}}}_{a1}}+\left( {{{\hat{H}}}_{0}}{{V}_{0}}+{{V}_{0}}{{{\hat{H}}}_{0}} \right)+{{V}_{0}}^{2} \right] \right\}.
\ee
Eliminate all terms that do not include ${{V}_{0}}$ from this expression to obtain
\be 
{{C}_{J1}}\left( {{V}_{0}} \right)\sim-\frac{1}{V{{M}^{3}}}T{{r}_{spin}}\int{D\bm{p}}\int{{{d}^{3}}x}\left\{ {{\sigma }_{z}}{{{{p}}}_{z}}^{{{n}_{z}}-1}\left[ \left( {{{\hat{H}}}_{0}}{{V}_{0}}+{{V}_{0}}{{{\hat{H}}}_{0}} \right)+{{V}_{0}}^{2} \right] \right\}.
\ee
Recall that since we are operating in the large $M$ limit the momentum operators in ${{\hat{H}}_{0}}$ are replaced by their corresponding eigenvalues. Taking the trace over spin gives
\be 
{{C}_{J1}}\left( {{V}_{0}} \right)\sim-\frac{4}{V{{M}^{3}}}\int{D\bm{p}}{{p}_{z}}^{2{{n}_{z}}-1}\int{{{d}^{3}}x}{{V}_{0}}=0.
\ee
This is zero because the integrand has an odd power of ${{p}_{z}}$.  
Next consider ${{C}_{J2}}$,
\be 
{{C}_{J2}}\sim \frac{1}{V{{M}^{5}}}T{{r}_{spin}}\left\{ \int{D\bm{p}}\int{{{d}^{3}}x}\left( {{\sigma }_{z}}{{p}_{z}}^{{{n}_{z}}-1}{{{\hat{H}}}_{b}}^{2} \right) \right\}.
\ee	
This becomes,
\be 
{{C}_{J2}}\sim \frac{1}{V{{M}^{5}}}T{{r}_{spin}}\left\{ \int{D\bm{p}}\int{{{d}^{3}}x}\,{{\sigma }_{z}}{{{\hat{p}}}_{z}}^{{{n}_{z}}-1}\left\{ \begin{aligned}
  & \left[ {{{\hat{H}}}_{a2}}+i{{\sigma }_{z}}{{{\hat{H}}}_{a1}}+\left\{ {{{\hat{H}}}_{0}},{{V}_{0}} \right\}_++{{V}_{0}}^{2} \right] \\ 
 & \times \left[ {{{\hat{H}}}_{a2}}+i{{\sigma }_{z}}{{{\hat{H}}}_{a1}}+\left\{ {{{\hat{H}}}_{0}},{{V}_{0}} \right\}_++{{V}_{0}}^{2} \right] \\ 
\end{aligned} \right\} \right\},
\ee
where $\left\{ \hat{A},\hat{B} \right\}_+=\hat{A}\hat{B}+\hat{B}\hat{A}$.
Next drop all terms with no dependence on ${{V}_{0}}$ to obtain
\be 
{{C}_{J2}}\left( {{V}_{0}} \right)\sim\frac{1}{V{{M}^{5}}}T{{r}_{spin}}\left\{ \int{d\bm{p}\int{{{d}^{3}}x}\,}{{\sigma }_{z}}{{p}_{z}}^{{{n}_{z}}-1}\left[ \begin{aligned}
  & \left\{ {{{\hat{H}}}_{a2}},\left\{ {{{\hat{H}}}_{0}},{{V}_{0}} \right\}_+ \right\}_++\left\{ i{{\sigma }_{z}}{{{\hat{H}}}_{a1}},\left\{ {{{\hat{H}}}_{0}},{{V}_{0}} \right\}_+ \right\}_++\left\{ i{{\sigma }_{z}}{{{\hat{H}}}_{a1}},V_{0}^{2} \right\}_+ \\ 
 & +\left\{ {{{\hat{H}}}_{a2}},V_{0}^{2} \right\}_++\left\{ {{{\hat{H}}}_{0}},{{V}_{0}} \right\}_+\left\{ {{{\hat{H}}}_{0}},{{V}_{0}} \right\}_++\left\{ \left\{ {{{\hat{H}}}_{0}},{{V}_{0}} \right\}_+,V_{0}^{2} \right\}_++V_{0}^{4} \\ 
\end{aligned} \right] \right\}.
\ee
We then use the properties of trace to eliminate some terms,
\be 
{{C}_{J2}}\left( {{V}_{0}} \right)\sim\frac{1}{{{M}^{5}}}T{{r}_{spin}}\left\{ \int{D\bm{p}}{{p}_{z}}^{{{n}_{z}}-1}\int{{{d}^{3}}x}\left[ \begin{aligned}
  & \left\{ {{{\hat{H}}}_{a2}},2{{p}_{z}}^{{{n}_{z}}}{{V}_{0}} \right\}_++i\left\{ {{{\hat{H}}}_{a1}},V_{0}^{2} \right\}_+ \\ 
 & +{{\sigma }_{z}}\left\{ {{{\hat{H}}}_{0}},{{V}_{0}} \right\}_+\left\{ {{{\hat{H}}}_{0}},{{V}_{0}} \right\}_++\left\{ 2{{p}_{z}}^{{{n}_{z}}}{{V}_{0}},V_{0}^{2} \right\}_+ \\ 
\end{aligned} \right] \right\}.
\ee	
Removing all terms that are odd in ${{{p}}_{z}}$, and noting the fact
\be 
T{{r}_{spin}}\left[{{\sigma }_{z}}\left\{ {{{\hat{H}}}_{0}},{{V}_{0}} \right\}_+\left\{ {{{\hat{H}}}_{0}},{{V}_{0}} \right\}_+\right]=T{{r}_{spin}}\left[ {{\sigma }_{z}}4{{{\hat{H}}}_{0}}^{2}{{V}_{0}}^{2} \right]=0,
\ee	
where we have used the fact that we can ignore commutators in the large $M$ limit,  we obtain
\be 
{{C}_{J2}}\left( {{V}_{0}} \right)\sim\frac{1}{V{{M}^{5}}}T{{r}_{spin}}\left\{ \int{D\bm{p}}\int{{{d}^{3}}x}\,{{p}_{z}}^{{{n}_{z}}-1}i2{{{\hat{H}}}_{a1}}V_{0}^{2} \right\}.
\ee	
The maximum order of the momentum in the term ${{\hat{H}}_{a1}}$ is ${{p}_{x}}^{{{n}_{x}}-1}{{p}_{y}}^{{{n}_{y}}-1}$.    Using this and Eq. (\ref{eqn:eqn040}) we show that
\be 
{{C}_{J2}}\left( {{V}_{0}} \right)\sim\frac{1}{V{{M}^{5}}}\left\{ \int D\bm{p}\,{{p}_{z}}^{{{n}_{z}}-1}{{p}_{x}}^{{{n}_{x}}-1}{{p}_{y}}^{{{n}_{y}}-1} \right\}\int{{{d}^{3}}x}V_{0}^{2}\sim\frac{{{M}^{3}}}{V{{M}^{5}}}\int{{{d}^{3}}x}V_{0}^{2} \to 0,
\ee	
in the limit $M\to \infty$, which shows that ${{C}_{J2}}$ has no dependence on ${{V}_{0}}\left( x,y,z \right)$.  The same argument applies to higher order terms, i.e., ${{C}_{Jm}}$, $m\ge 3$, which proves that ${{C}_{J}}$ is not dependent on ${{V}_{0}}\left( x,y,z \right)$.

\begin{center}
{\bf Section 4}  
\end{center}

 In the previous sections we have presented proofs that $C_J$ is independent of a perturbing potential for a variety of different Hamiltonians. An intriguing fact is that in order for this to be true it requires precise cancellations of all terms that are dependent on the perturbation. To demonstrate this assertion we will do a detailed calculation to show that ${{C}_{J}}$ is independent of the perturbing potential $\hat{V}\left( x,y \right)$ up to the second order in $\hat{V}$ for arbitrary $M>0$ for the following Hamiltonian,
\be 
{{\hat{H}}_{1a\lambda }}={{\hat{H}}_{0}}+ \hat{V}\left( x,y \right),
\ee	
where ${{\hat{H}}_{0}}=\bm{\sigma }\cdot \bm{\pi }$ and $\hat{V}\left( x,y \right)={{V}_{0}}\left( x,y \right)+\bm{\sigma }\cdot \bm{V}\left( x,y \right)$.  
Since we intend to show that ${{C}_{J}}$ is independent of $\hat{V}\left( x,y \right)$ up to the second order in $\hat{V}$ we will only retain terms up to the second order in $\hat{V}$ in the following discussion.
	 
We will use the following relations.   First, define ${{\hat{H}}_{a}}={{\sigma }_{x}}{{\hat{p}}_{x}}+{{\sigma }_{y}}\left( {{{\hat{p}}}_{y}}-Bx \right)$ which gives ${{\hat{H}}_{0}}={{\hat{H}}_{a}}+{{\sigma }_{z}}{{\hat{p}}_{z}}$.  Next write
\be {{\left( {{{\hat{H}}}_{0}}+\hat{V} \right)}^{2}}={{\hat{H}}_{0}}^{2}+{{\hat{H}}_{1}},\quad {{\hat{H}}_{1}}=\left( {{{\hat{H}}}_{0}}\hat{V}+\hat{V}{{{\hat{H}}}_{0}} \right)+{{\hat{V}}^{2}},\quad {{\hat{H}}_{0}}^{2}=\hat{H}_{a}^{2}+\hat{p}_{z}^{2}.
\ee
From the above we also have
\be \label{eqn:eq12}
{{\hat{H}}_{1}}=\left[ \left( {{{\hat{H}}}_{a}}+{{\sigma }_{z}}{{{\hat{p}}}_{z}} \right)\hat{V}+\hat{V}\left( {{{\hat{H}}}_{a}}+{{\sigma }_{z}}{{{\hat{p}}}_{z}} \right) \right]+{{\hat{V}}^{2}}.
\ee	
From all this we obtain,
\be {{e}^{-\frac{\hat{H}_{1a\lambda }^{2}}{{{M}^{2}}}}}={{e}^{-\frac{{{\left( {{{\hat{H}}}_{0}}+\hat{V} \right)}^{2}}}{{{M}^{2}}}}}={{e}^{-\frac{\left( \hat{H}_{a}^{2}+\hat{p}_{z}^{2} \right)+{{{\hat{H}}}_{1}}}{{{M}^{2}}}}}={{e}^{-\frac{\hat{p}_{z}^{2}}{{{M}^{2}}}}}{{e}^{-\frac{\hat{H}_{a}^{2}+{{{\hat{H}}}_{1}}}{{{M}^{2}}}}},
\ee	
where we have used the fact that ${{\hat{p}}_{z}}$ commutes with ${{\hat{H}}_{a}}$ and ${{\hat{H}}_{1}}$.  Next we use the perturbation theory to have		
\be 
{{e}^{-\frac{\hat{H}_{a}^{2}+{{{\hat{H}}}_{1}}}{{{M}^{2}}}}}={{e}^{-\frac{\hat{H}_{a}^{2}}{{{M}^{2}}}}}\left\{ 1-\frac{1}{{{M}^{2}}}\int\limits_{0}^{1}{{{{\hat{H}}}_{1}}\left( t \right)dt}+\frac{1}{{{M}^{4}}}\int\limits_{0}^{1}{{{{\hat{H}}}_{1}}\left( t \right)dt}\int\limits_{0}^{t}{{{{\hat{H}}}_{1}}\left( {{t}'} \right)d{t}'}+O\left( {{V}^{3}} \right) \right\},
\ee	
where
\be 
{{\hat{H}}_{1}}\left( t \right)={{e}^{+\frac{\hat{H}_{a}^{2}}{{{M}^{2}}}t}}{{\hat{H}}_{1}}{{e}^{-\frac{\hat{H}_{a}^{2}}{{{M}^{2}}}t}}=\left[ \left( {{{\hat{H}}}_{a}}+{{\sigma }_{z}}{{{\hat{p}}}_{z}} \right)\hat{V}\left( t \right)+\hat{V}\left( t \right)\left( {{{\hat{H}}}_{a}}+{{\sigma }_{z}}{{{\hat{p}}}_{z}} \right) \right]+\hat{V}{{\left( t \right)}^{2}},\quad
\hat{V}\left( t \right)={{e}^{+\frac{\hat{H}_{a}^{2}}{{{M}^{2}}}t}}\hat{V}{{e}^{-\frac{\hat{H}_{a}^{2}}{{{M}^{2}}}t}}.
\ee	
We can use the above results to show that ${{C}_{J}}$ can be expanded as ${{C}_{J}}={{C}_{J0}}+{{C}_{J1}}+{{C}_{J2}}+\ldots $ in powers of $\hat V$.  Since we are interested in the dependence on $\hat{V}$ up to the second order, the only terms in this expansion that we need to analyze are ${{C}_{J1}}$ and ${{C}_{J2}}$, which are given by
\be \label{eqn:eq17}
{{C}_{J1}}=\frac{-1}{{{M}^{3}}V\sqrt{\pi }}Tr\left\{ {{\sigma }_{z}}{{e}^{-\frac{\hat{p}_{z}^{2}}{{{M}^{2}}}}}{{e}^{-\frac{\hat{H}_{a}^{2}}{{{M}^{2}}}}}\int\limits_{0}^{1}{{{{\hat{H}}}_{1}}\left( t \right)dt} \right\},
\ee	
and,
\be \label{eqn:eq060}
{{C}_{J2}}=\frac{1}{{{M}^{5}}V\sqrt{\pi }}Tr\left\{ {{\sigma }_{z}}{{e}^{-\frac{\hat{p}_{z}^{2}}{{{M}^{2}}}}}{{e}^{-\frac{\hat{H}_{a}^{2}}{{{M}^{2}}}}}\int\limits_{0}^{1}{{{{\hat{H}}}_{1}}\left( t \right)dt}\int\limits_{0}^{t}{{{{\hat{H}}}_{1}}\left( {{t}'} \right)d{t}'} \right\}.
\ee	
Let us first evaluate ${{C}_{J1}}$.  Use Eq. (\ref{eqn:eq12}) in Eq. (\ref{eqn:eq17}) to obtain
\be 
{{C}_{J1}}=\frac{-1}{{{M}^{3}}V\sqrt{\pi }}Tr\left\{ {{\sigma }_{z}}{{e}^{-\frac{\hat{p}_{z}^{2}}{{{M}^{2}}}}}{{e}^{-\frac{\hat{H}_{a}^{2}}{{{M}^{2}}}}}\int\limits_{0}^{1}{\left[ \left( {{{\hat{H}}}_{a}}+{{\sigma }_{z}}{{{\hat{p}}}_{z}} \right)\hat{V}\left( t \right)+\hat{V}\left( t \right)\left( {{{\hat{H}}}_{a}}+{{\sigma }_{z}}{{{\hat{p}}}_{z}} \right) \right]+\hat{V}{{\left( t \right)}^{2}}dt} \right\}.
\ee		
Next take the trace over ${{\hat{p}}_{z}}$ and use $\sigma_z\hat{H}_a=-\hat{H}_a\sigma_z$ to remove all terms that are first order in ${{\hat{H}}_{a}}$ and ${{\hat{p}}_{z}}$ to obtain
\be 
 {{C}_{J1}}=\frac{-1}{{{M}^{3}}V\sqrt{\pi }}\int\limits_{-\infty }^{+\infty }{\frac{{{L}_{z}}d{{p}_{z}}}{2\pi }{{e}^{-\frac{p_{z}^{2}}{{{M}^{2}}}}}}Tr\left\{ {{\sigma }_{z}}{{e}^{-\frac{\hat{H}_{a}^{2}}{{{M}^{2}}}}}\int\limits_{0}^{1}{\hat{V}{{\left( t \right)}^{2}}dt} \right\}. \label{eqn:eq20}
\ee
The eigenfunctions and eigenvalues of ${{\hat{H}}_{a}}$ are the well-known solutions of the two dimensional relativistic Landau level problem and are given as
\be 
{{\hat{H}}_{a}}\left| {{\psi }_{n,{{p}_{y}},{{\beta }_{n}}}} \right\rangle ={{\beta }_{n}}{{\lambda }_{n}}\left| {{\psi }_{n,{{p}_{y}},{{\beta }_{n}}}} \right\rangle ,\quad {{\lambda }_{n}}=\sqrt{2Bn},\quad \left| {{\psi }_{_{n,{{p}_{y}},{{\beta }_{n}}}}} \right\rangle =\frac{1}{\sqrt{V}}{{e}^{i{{p}_{y}}y}}{{Y}_{n,{{p}_{y}},{{\beta }_{n}}}}\left( x \right),\quad n=0,1,2,\ldots,
\ee		
where ${{\beta }_{n}}=\pm 1$ for $n>0$ and ${{\beta }_{0}}=0$ for $n=0$ and we assume $B>0$.  It can also be shown from $\sigma_z{\hat H}_a=-\hat{H}_a\sigma_z$ that ${{\sigma }_{z}}\left| {{\psi }_{n,{{p}_{y}},-{{\beta }_{n}}}} \right\rangle $ is an energy eigenfunction with the energy ${{\beta }_{n}}{{\lambda }_{n}}$.  Therefore,
\be
\langle {{\psi }_{n,{{p}_{y}},{{\beta }_{n}}}}|{{\sigma }_{z}}\left| {{\psi }_{m,{{q}_{y}},{{\alpha }_{m}}}} \right\rangle \propto {{\delta }_{n,m}}{{\delta }_{{{p}_{y}},{{q}_{y}}}}{{\delta }_{{{\beta }_{n}},-{{\alpha }_{m}}}}.
\ee	
Use this in Eq. (\ref{eqn:eq20}) to obtain
\be 
{{C}_{J1}}=\frac{-1}{{{M}^{3}}V\sqrt{\pi }}\int\limits_{-\infty }^{+\infty }{\frac{{{L}_{z}}d{{p}_{z}}}{2\pi }{{e}^{-\frac{p_{z}^{2}}{{{M}^{2}}}}}}\sum\limits_{n,{{p}_{y}},{{\beta }_{n}}}{\left\{ {{e}^{-\frac{\lambda _{n}^{2}}{{{M}^{2}}}}}\left\langle  {{\psi }_{n,{{p}_{y}},{{\beta }_{n}}}} \right|{{\sigma }_{z}}\left| {{\psi }_{n,{{p}_{y}},-{{\beta }_{n}}}} \right\rangle \left\langle  {{\psi }_{n,{{p}_{y}},-{{\beta }_{n}}}} \right|{{{\hat{V}}}^{2}}\left| {{\psi }_{n,{{p}_{y}},{{\beta }_{n}}}} \right\rangle  \right\}},
\ee	
where we have used $\langle {{\psi }_{n,{{p}_{y}},-{{\beta }_{n}}}}|\hat{V}{{\left( t \right)}^{2}}\left| {{\psi }_{n,{{p}_{y}},{{\beta }_{n}}}} \right\rangle =\langle {{\psi }_{n,{{p}_{y}},-{{\beta }_{n}}}}|{{\hat{V}}^{2}}\left| {{\psi }_{n,{{p}_{y}},{{\beta }_{n}}}} \right\rangle $.  Evaluating the $p_z$-integral, we obtain
\be 
{{C}_{J1}}=\frac{-{{L}_{z}}}{2\pi {{M}^{2}}V}\sum\limits_{n,{{p}_{y}},\beta_n }{\left\{ {{e}^{-\frac{\lambda _{n}^{2}}{{{M}^{2}}}}}\left\langle  {{\psi }_{n,{{p}_{y}},\beta_n }} \right|{{\sigma }_{z}}\left| {{\psi }_{n,{{p}_{y}},-\beta_n }} \right\rangle \left\langle  {{\psi }_{n,{{p}_{y}},-\beta_n }} \right|{{{\hat{V}}}^{2}}\left| {{\psi }_{n,{{p}_{y}},\beta_n }} \right\rangle  \right\}}.
\ee
Note that we can write the last term as
\be 
\left\langle  {{\psi }_{n,{{p}_{y}},-{{\beta }_{n}}}} \right|{{\hat{V}}^{2}}\left| {{\psi }_{n,{{p}_{y}},{{\beta }_{n}}}} \right\rangle =\sum\limits_{m,{{q}_{y}},\alpha_m }{\left\langle  {{\psi }_{n,{{p}_{y}},-{{\beta }_{n}}}} \right|\hat{V}\left| {{\psi }_{m,{{q}_{y}},{{\alpha }_{m}}}} \right\rangle \left\langle  {{\psi }_{m,{{q}_{y}},{{\alpha }_{m}}}} \right|\hat{V}\left| {{\psi }_{n,{{p}_{y}},{{\beta }_{n}}}} \right\rangle },
\ee	
which gives the expression
\be \label{eqn:eq26}
{{C}_{J1}}=\frac{-{{L}_{z}}}{2\pi {{M}^{2}}V}\sum\limits_{n,{{p}_{y}},{{\beta }_{n}}}{\sum\limits_{m,{{q}_{y,}}{{\alpha }_{m}}}{\left\{ {{e}^{-\frac{\lambda _{n}^{2}}{{{M}^{2}}}}}\left\langle  {{\psi }_{n,{{p}_{y}},{{\beta }_{n}}}} \right|{{\sigma }_{z}}\left| {{\psi }_{n,{{p}_{y}},-{{\beta }_{n}}}} \right\rangle \left\langle  {{\psi }_{n,{{p}_{y}},-{{\beta }_{n}}}} \right|\hat{V}\left| {{\psi }_{m,{{q}_{y}},{{\alpha }_{m}}}} \right\rangle \left\langle  {{\psi }_{m,{{q}_{y}},{{\alpha }_{m}}}} \right|\hat{V}\left| {{\psi }_{n,{{p}_{y}},{{\beta }_{n}}}} \right\rangle  \right\}}}.
\ee	
Next we will evaluate ${{C}_{J2}}$.  Refer to Eq. (\ref{eqn:eq060}) and take the trace over ${{\hat{p}}_{z}}$ to obtain 
\be \label{eqn:eq27}
{{C}_{J2}}=\frac{1}{{{M}^{5}}V\sqrt{\pi }}\int\limits_{-\infty }^{+\infty }{\frac{{{L}_{z}}d{{p}_{z}}}{2\pi }{{e}^{-\frac{p_{z}^{2}}{{{M}^{2}}}}}}Tr\left\{ {{\sigma }_{z}}{{e}^{-\frac{\hat{H}_{a}^{2}}{{{M}^{2}}}}}\int\limits_{0}^{1}{dt}\int\limits_{0}^{t}d{t}'{{{{\hat{H}}}_{1}}\left( t \right){{{\hat{H}}}_{1}}\left( {{t}'} \right)} \right\},
\ee
where,
\be 
{{\hat{H}}_{1}}\left( t \right){{\hat{H}}_{1}}\left( {{t}'} \right)=\left( \begin{aligned}
  & \left( {{{\hat{H}}}_{a}}+{{\sigma }_{z}}{{p}_{z}} \right)\hat{V}\left( t \right) \\ 
 & +\hat{V}\left( t \right)\left( {{{\hat{H}}}_{a}}+{{\sigma }_{z}}{{p}_{z}} \right)+\hat{V}{{\left( t \right)}^{2}} \\ 
\end{aligned} \right)\left( \begin{aligned}
  & \left( {{{\hat{H}}}_{a}}+{{\sigma }_{z}}{{p}_{z}} \right)\hat{V}\left( {{t}'} \right) \\ 
 & +\hat{V}\left( {{t}'} \right)\left( {{{\hat{H}}}_{a}}+{{\sigma }_{z}}{{p}_{z}} \right)+\hat{V}{{\left( {{t}'} \right)}^{2}} \\ 
\end{aligned} \right).
\ee	
Rewrite this and only retain terms to the second order in $\hat{V}$ to arrive at
\be 
{{\hat{H}}_{1}}\left( t \right){{\hat{H}}_{1}}\left( {{t}'} \right)\to \left( \begin{aligned}
  & \left( {{{\hat{H}}}_{a}}+{{\sigma }_{z}}{{{{p}}}_{z}} \right)\hat{V}\left( t \right) \\ 
 & +\hat{V}\left( t \right)\left( {{{\hat{H}}}_{a}}+{{\sigma }_{z}}{{{{p}}}_{z}} \right) \\ 
\end{aligned} \right)\left( \begin{aligned}
  & \left( {{{\hat{H}}}_{a}}+{{\sigma }_{z}}{{{{p}}}_{z}} \right)\hat{V}\left( {{t}'} \right) \\ 
 & +\hat{V}\left( {{t}'} \right)\left( {{{\hat{H}}}_{a}}+{{\sigma }_{z}}{{{{p}}}_{z}} \right) \\ 
\end{aligned} \right).
\ee	
Next use the fact that the terms to odd powers in ${{p}_{z}}$ drop out after integration to obtain,
\be 
{{\hat{H}}_{1}}\left( t \right){{\hat{H}}_{1}}\left( {{t}'} \right)\to \left[ \begin{aligned}
  & \left( {{{\hat{H}}}_{a}}\hat{V}\left( t \right){{{\hat{H}}}_{a}}\hat{V}\left( {{t}'} \right)+\hat{V}\left( t \right){{{\hat{H}}}_{a}}\hat{V}\left( {{t}'} \right){{{\hat{H}}}_{a}} \right) \\ 
 & +\left( {{{\hat{H}}}_{a}}\hat{V}\left( t \right)\hat{V}\left( {{t}'} \right){{{\hat{H}}}_{a}}+\hat{V}\left( t \right){{{\hat{H}}}_{a}}^{2}\hat{V}\left( {{t}'} \right) \right) \\ 
\end{aligned} \right]+p_{z}^{2}\left[ \begin{aligned}
  & {{\sigma }_{z}}\hat{V}\left( t \right){{\sigma }_{z}}\hat{V}\left( {{t}'} \right)+{{\sigma }_{z}}\hat{V}\left( t \right)\hat{V}\left( {{t}'} \right){{\sigma }_{z}} \\ 
 & +\hat{V}\left( t \right)\hat{V}\left( {{t}'} \right)+\hat{V}\left( t \right){{\sigma }_{z}}\hat{V}\left( {{t}'} \right){{\sigma }_{z}} \\ 
\end{aligned} \right].
\ee		
When this term is substituted into Eq. (\ref{eqn:eq27}) it can be shown, using the properties of the trace and the fact that ${{\hat{H}}_{a}}$ anticommutes with ${{\sigma }_{z}}$, that the above relation can be replaced by
\be 
{{\sigma }_{z}}{{\hat{H}}_{1}}\left( t \right){{\hat{H}}_{1}}\left( {{t}'} \right)\to {{\sigma }_{z}}\left( -\left[ {{{\hat{H}}}_{a}}^{2},\hat{V}\left( t \right) \right]\hat{V}\left( {{t}'} \right) \right)+2p_{z}^{2}\left[ \hat{V}\left( t \right){{\sigma }_{z}}\hat{V}\left( {{t}'} \right)+{{\sigma }_{z}}\hat{V}\left( t \right)\hat{V}\left( {{t}'} \right) \right].
\ee	
Use this result in (\ref{eqn:eq27}) to obtain ${{C}_{J2}}={{C}_{J2a}}+{{C}_{J2b}}$ where,
\be 
{{C}_{J2a}}=\frac{1}{{{M}^{5}}V\sqrt{\pi }}\int\limits_{-\infty }^{+\infty }{\frac{{{L}_{z}}d{{p}_{z}}}{2\pi }{{e}^{-\frac{p_{z}^{2}}{{{M}^{2}}}}}}Tr\left\{ {{\sigma }_{z}}{{e}^{-\frac{\hat{H}_{a}^{2}}{{{M}^{2}}}}}\int\limits_{0}^{1}{dt}\int\limits_{0}^{t}d{t}'{\left( -\left[ {{{\hat{H}}}_{a}}^{2},\hat{V}\left( t \right) \right]\hat{V}\left( {{t}'} \right) \right)} \right\},
\ee		
and, 
\be {{C}_{J2b}}=\frac{2}{{{M}^{5}}V\sqrt{\pi }}\int\limits_{-\infty }^{+\infty }{\frac{{{L}_{z}}d{{p}_{z}}}{2\pi }p_{z}^{2}{{e}^{-\frac{p_{z}^{2}}{{{M}^{2}}}}}}Tr\left\{ {{e}^{-\frac{\hat{H}_{a}^{2}}{{{M}^{2}}}}}\int\limits_{0}^{1}{dt}\int\limits_{0}^{t}d{t}'{\left[ \hat{V}\left( t \right){{\sigma }_{z}}\hat{V}\left( {{t}'} \right)+{{\sigma }_{z}}\hat{V}\left( t \right)\hat{V}\left( {{t}'} \right) \right]} \right\}.
\ee	
Recall that in the above expressions we have dropped terms of higher order than  ${{\hat{V}}^{2}}$.
Evaluating the trace, we have
\be \label{eqn:eq34}
{{C}_{J2a}}=\frac{1}{{{M}^{5}}V\sqrt{\pi }}\int\limits_{-\infty }^{+\infty }{\frac{{{L}_{z}}d{{p}_{z}}}{2\pi }{{e}^{-\frac{p_{z}^{2}}{{{M}^{2}}}}}}\sum\limits_{n,{{p}_{y}},{{\beta }_{n}}}{{{e}^{-\frac{\lambda _{n}^{2}}{{{M}^{2}}}}}}\left\{ \begin{aligned}
  & \left\langle  {{\psi }_{n,{{p}_{y}},{{\beta }_{n}}}} \right|{{\sigma }_{z}}\left| {{\psi }_{n,{{p}_{y}},-{{\beta }_{n}}}} \right\rangle \left\langle  {{\psi }_{n,{{p}_{y}},-{{\beta }_{n}}}} \right| \\ 
 & \times \int\limits_{0}^{1}{dt}\int\limits_{0}^{t}d{t}'{\left( -\left[ {{{\hat{H}}}_{a}}^{2},\hat{V}\left( t \right) \right]\hat{V}\left( {{t}'} \right) \right)\left| {{\psi }_{n,{{p}_{y}},{{\beta }_{n}}}} \right\rangle } \\ 
\end{aligned} \right\},
\ee		
and,
\be \label{eqn:eq35}
{{C}_{J2b}}=\frac{2}{{{M}^{5}}V\sqrt{\pi }}\int\limits_{-\infty }^{+\infty }{\frac{{{L}_{z}}d{{p}_{z}}}{2\pi }p_{z}^{2}{{e}^{-\frac{p_{z}^{2}}{{{M}^{2}}}}}}\sum\limits_{n,{{p}_{y}},{{\beta }_{n}}}{\left\{ {{e}^{-\frac{\lambda _{n}^{2}}{{{M}^{2}}}}}\int\limits_{0}^{1}{dt}\int\limits_{0}^{t}d{t}'{\left\langle  {{\psi }_{n,{{p}_{y}},{{\beta }_{n}}}} \right|\left[ \begin{aligned}
  & \hat{V}\left( t \right){{\sigma }_{z}}\hat{V}\left( {{t}'} \right) \\ 
 & +{{\sigma }_{z}}\hat{V}\left( t \right)\hat{V}\left( {{t}'} \right) \\ 
\end{aligned} \right]\left| {{\psi }_{n,{{p}_{y}},{{\beta }_{n}}}} \right\rangle } \right\}}.
\ee		
We first evaluate ${{C}_{J2a}}$.  We have
\be \left\langle  {{\psi }_{n,{{p}_{y}},-{{\beta }_{n}}}} \right|\hat{V}\left( t \right)\hat{V}\left( {{t}'} \right)\left| {{\psi }_{n,{{p}_{y}},{{\beta }_{n}}}} \right\rangle =\sum\limits_{m,{{q}_{y}},\alpha_m }{\left\langle  {{\psi }_{n,{{p}_{y}},-{{\beta }_{n}}}} \right|\hat{V}\left| {{\psi }_{m,{{q}_{y}},{{\alpha }_{m}}}} \right\rangle \left\langle  {{\psi }_{m,{{q}_{y}},{{\alpha }_{m}}}} \right|\hat{V}\left| {{\psi }_{n,{{p}_{y}},{{\beta }_{n}}}} \right\rangle }{{e}^{+\frac{\lambda _{n}^{2}-\lambda _{m}^{2}}{{{M}^{2}}}t}}{{e}^{+\frac{\lambda _{m}^{2}-\lambda _{n}^{2}}{{{M}^{2}}}{t}'}}.
\ee		
Using the integral
\be \int\limits_{0}^{1}{dt}\int\limits_{0}^{t}d{t}'{{{e}^{+\frac{{{\lambda }_{n}^{2}}-{{\lambda }_{m}^{2}}}{{{M}^{2}}}t}}{{e}^{+\frac{{{\lambda }_{m}^{2}}-{{\lambda }_{n}^{2}}}{{{M}^{2}}}{t}'}}}=\left[ \frac{{{M}^{2}}}{\lambda _{m}^{2}-\lambda _{n}^{2}}+\frac{{{M}^{4}}\left( {{e}^{+\frac{\lambda _{n}^{2}-\lambda _{m}^{2}}{{{M}^{2}}}}}-1 \right)}{{{\left( \lambda _{m}^{2}-\lambda _{n}^{2} \right)}^{2}}} \right],
\ee
we can show that $\int\limits_{0}^{1}{dt}\int\limits_{0}^{t}d{t}'{\langle {{\psi }_{n,{{p}_{y}},-{{\beta }_{n}}}}|\left[ -\left[ {{{\hat{H}}}_{a}}^{2},\hat{V}\left( t \right) \right]\hat{V}\left( {{t}'} \right) \right]\left| {{\psi }_{n,{{p}_{y}}.{{\beta }_{n}}}} \right\rangle}$ is given by,
\be \sum\limits_{m,{{q}_{y}},{{\alpha }_{m}}}{\left\langle  {{\psi }_{n,{{p}_{y}}-{{\beta }_{n}}}} \right|\hat{V}\left| {{\psi }_{m,{{q}_{y}},{{\alpha }_{m}}}} \right\rangle \left\langle  {{\psi }_{m,{{q}_{y}},{{\alpha }_{m}}}} \right|\hat{V}\left| {{\psi }_{n,{{p}_{y}},{{\beta }_{n}}}} \right\rangle }\left[ -\left( \lambda _{n}^{2}-\lambda _{m}^{2} \right) \right]\left[ \frac{{{M}^{2}}}{\lambda _{m}^{2}-\lambda _{n}^{2}}+\frac{{{M}^{4}}\left( {{e}^{+\frac{\lambda _{n}^{2}-\lambda _{m}^{2}}{{{M}^{2}}}}}-1 \right)}{{{\left( \lambda _{m}^{2}-\lambda _{n}^{2} \right)}^{2}}} \right].
\ee	
Use this result in Eq. (\ref{eqn:eq34}) and performing the Gaussian integration to obtain
\be 
{{C}_{J2a}}=\frac{{{L}_{z}}}{2\pi {{M}^{2}}V}\sum\limits_{n,{{p}_{y}},{{\beta }_{n}}}{\sum\limits_{m,{{q}_{y}},{{\alpha }_{m}}}{\left\{ \begin{aligned}
  & \left\langle  {{\psi }_{n,{{p}_{y}},{{\beta }_{n}}}} \right|{{\sigma }_{z}}\left| {{\psi }_{n,{{p}_{y}},-{{\beta }_{n}}}} \right\rangle \left\langle  {{\psi }_{n,{{p}_{y}}-{{\beta }_{n}}}} \right|\hat{V}\left| {{\psi }_{m,{{q}_{y}},{{\alpha }_{m}}}} \right\rangle  \\ 
 & \times \left\langle  {{\psi }_{m,{{q}_{y}},{{\alpha }_{m}}}} \right|\hat{V}\left| {{\psi }_{n,{{p}_{y}},{{\beta }_{n}}}} \right\rangle \left[ {{e}^{-\frac{\lambda _{n}^{2}}{{{M}^{2}}}}}-\frac{{{M}^{2}}\left( {{e}^{-\frac{\lambda _{m}^{2}}{{{M}^{2}}}}}-{{e}^{-\frac{\lambda _{n}^{2}}{{{M}^{2}}}}} \right)}{\left( \lambda _{n}^{2}-\lambda _{m}^{2} \right)} \right] \\ 
\end{aligned} \right\}}}.
\ee	
Adding  ${{C}_{J1}}$ (see Eq. (\ref{eqn:eq26})) to this expression, we obtain
\be \label{eqn:eq40}
{{C}_{J2a}}+{{C}_{J1}}=\frac{-{{L}_{z}}}{2\pi V}\sum\limits_{n,{{p}_{y}},{{\beta }_{n}}}{\sum\limits_{m,{{q}_{y}},{{\alpha }_{m}}}{\left\{ \begin{aligned}
  & \left\langle  {{\psi }_{n,{{p}_{y}},{{\beta }_{n}}}} \right|{{\sigma }_{z}}\left| {{\psi }_{n,{{p}_{y}},-{{\beta }_{n}}}} \right\rangle \left\langle  {{\psi }_{n,{{p}_{y}}-{{\beta }_{n}}}} \right|\hat{V}\left| {{\psi }_{m,{{q}_{y}},{{\alpha }_{m}}}} \right\rangle  \\ 
 & \times \left\langle  {{\psi }_{m,{{q}_{y}},{{\alpha }_{m}}}} \right|\hat{V}\left| {{\psi }_{n,{{p}_{y}},{{\beta }_{n}}}} \right\rangle \left[ \frac{\left( {{e}^{-\frac{\lambda _{m}^{2}}{{{M}^{2}}}}}-{{e}^{-\frac{\lambda _{n}^{2}}{{{M}^{2}}}}} \right)}{\left( \lambda _{n}^{2}-\lambda _{m}^{2} \right)} \right] \\ 
\end{aligned} \right\}}}.
\ee	
We next calculate ${{C}_{J2b}}$.  First note that
\be 
\left\langle  {{\psi }_{n,{{p}_{y}},{{\beta }_{n}}}} \right|\hat{V}\left( t \right){{\sigma }_{z}}\hat{V}\left( {{t}'} \right)\left| {{\psi }_{n,{{p}_{y}},{{\beta }_{n}}}} \right\rangle =\sum\limits_{m,{{q}_{y}},{{\alpha }_{m}}}{\left\{ \begin{aligned}
  & \left\langle  {{\psi }_{n,{{p}_{y}},{{\beta }_{n}}}} \right|\hat{V}\left| {{\psi }_{m,{{q}_{y}},{{\alpha }_{m}}}} \right\rangle  \\ 
 & \times \left\langle  {{\psi }_{m,{{q}_{y}},{{\alpha }_{m}}}} \right|{{\sigma }_{z}}\left| {{\psi }_{m,{{q}_{y}},-{{\alpha }_{m}}}} \right\rangle  \\ 
 & \times \left\langle  {{\psi }_{m,{{q}_{y}},-{{\alpha }_{m}}}} \right|\hat{V}\left| {{\psi }_{n,{{p}_{y}},{{\beta }_{n}}}} \right\rangle  \\ 
\end{aligned} \right\}{{e}^{+\frac{\lambda _{n}^{2}-\lambda _{m}^{2}}{{{M}^{2}}}t}}{{e}^{+\frac{\lambda _{m}^{2}-\lambda _{n}^{2}}{{{M}^{2}}}{t}'}}},
\ee		
and
\be 
\left\langle  {{\psi }_{n,{{p}_{y}},{{\beta }_{n}}}} \right|{{\sigma }_{z}}\hat{V}\left( t \right)\hat{V}\left( {{t}'} \right)\left| {{\psi }_{n,{{p}_{y}},{{\beta }_{n}}}} \right\rangle =\sum\limits_{m,{{q}_{y}},{{\alpha }_{m}}}{\left\{ \begin{aligned}
  & \left\langle  {{\psi }_{n,{{p}_{y}},{{\beta }_{n}}}} \right|{{\sigma }_{z}}\left| {{\psi }_{n,{{p}_{y}},-{{\beta }_{n}}}} \right\rangle  \\ 
 & \times \left\langle  {{\psi }_{n,{{p}_{y}},-{{\beta }_{n}}}} \right|\hat{V}\left| {{\psi }_{m,{{q}_{y}},{{\alpha }_{m}}}} \right\rangle  \\ 
 & \times \left\langle  {{\psi }_{m,{{q}_{y}},{{\alpha }_{m}}}} \right|\hat{V}\left| {{\psi }_{n,{{p}_{y}},{{\beta }_{n}}}} \right\rangle  \\ 
\end{aligned} \right\}}{{e}^{+\frac{\lambda _{n}^{2}-\lambda _{m}^{2}}{{{M}^{2}}}t}}{{e}^{+\frac{\lambda _{m}^{2}-\lambda _{n}^{2}}{{{M}^{2}}}{t}'}}.
\ee
Use these results in Eq. (\ref{eqn:eq35}) to obtain ${{C}_{J2b}}={{C}_{J2bA}}+{{C}_{J2bB}}$, where
\be \label{eqn:eq43}
{{C}_{J2bA}}=\frac{{{L}_{z}}}{2\pi V}\left\{ \sum\limits_{n,{{p}_{y}},{{\beta }_{n}}}{\sum\limits_{m,{{q}_{y}},{{\alpha }_{m}}}{\left\{ \begin{aligned}
  & \left\langle  {{\psi }_{n,{{p}_{y}},{{\beta }_{n}}}} \right|\hat{V}\left| {{\psi }_{m,{{q}_{y}},{{\alpha }_{m}}}} \right\rangle  \\ 
 & \times \left\langle  {{\psi }_{m,{{q}_{y}},{{\alpha }_{m}}}} \right|{{\sigma }_{z}}\left| {{\psi }_{m,{{q}_{y}},-{{\alpha }_{m}}}} \right\rangle  \\ 
 & \times \left\langle  {{\psi }_{m,{{q}_{y}},-{{\alpha }_{m}}}} \right|\hat{V}\left| {{\psi }_{n,{{p}_{y}},{{\beta }_{n}}}} \right\rangle  \\ 
\end{aligned} \right\}\left[ \frac{{{e}^{-\frac{\lambda _{n}^{2}}{{{M}^{2}}}}}}{\lambda _{m}^{2}-\lambda _{n}^{2}}+\frac{{{M}^{2}}\left( {{e}^{-\frac{\lambda _{m}^{2}}{{{M}^{2}}}}}-{{e}^{-\frac{\lambda _{n}^{2}}{{{M}^{2}}}}} \right)}{{{\left( \lambda _{m}^{2}-\lambda _{n}^{2} \right)}^{2}}} \right]}} \right\},
\ee		
and,
\be \label{eqn:eq44}
{{C}_{J2bB}}=\frac{{{L}_{z}}}{2\pi V}\left\{ \sum\limits_{n,{{p}_{y}},{{\beta }_{n}}}{\sum\limits_{m,{{q}_{y}},{{\alpha }_{m}}}{\left\{ \begin{aligned}
  & \left\langle  {{\psi }_{n,{{p}_{y}},{{\beta }_{n}}}} \right|{{\sigma }_{z}}\left| {{\psi }_{n,{{p}_{y}},-{{\beta }_{n}}}} \right\rangle  \\ 
 & \times \left\langle  {{\psi }_{n,{{p}_{y}},-{{\beta }_{n}}}} \right|\hat{V}\left| {{\psi }_{m,{{q}_{y}},{{\alpha }_{m}}}} \right\rangle  \\ 
 & \times \left\langle  {{\psi }_{m,{{q}_{y}},{{\alpha }_{m}}}} \right|\hat{V}\left| {{\psi }_{n,{{p}_{y}},{{\beta }_{n}}}} \right\rangle  \\ 
\end{aligned} \right\}\left[ \frac{{{e}^{-\frac{\lambda _{n}^{2}}{{{M}^{2}}}}}}{\lambda _{m}^{2}-\lambda _{n}^{2}}+\frac{{{M}^{2}}\left( {{e}^{-\frac{\lambda _{m}^{2}}{{{M}^{2}}}}}-{{e}^{-\frac{\lambda _{n}^{2}}{{{M}^{2}}}}} \right)}{{{\left( \lambda _{m}^{2}-\lambda _{n}^{2} \right)}^{2}}} \right]}} \right\}.
\ee		
We switch the dummy indices $n,{{p}_{y}},{{\beta }_{n}}$ and $m,{{q}_{y}},{{\alpha }_{m}}$ in Eq. (\ref{eqn:eq43}) to rewrite it as
\be 
{{C}_{J2bA}}=\frac{{{L}_{z}}}{2\pi V}\left\{ \sum\limits_{n,{{p}_{y}},{{\beta }_{n}}}{\sum\limits_{m,{{q}_{y}},{{\alpha }_{m}}}{\left\{ \begin{aligned}
  & \left\langle  {{\psi }_{m,{{q}_{y}},{{a}_{m}}}} \right|\hat{V}\left| {{\psi }_{n,{{p}_{y}},{{\beta }_{n}}}} \right\rangle  \\ 
 & \times \left\langle  {{\psi }_{n,{{p}_{y}},{{\beta }_{n}}}} \right|{{\sigma }_{z}}\left| {{\psi }_{n,{{p}_{y}},-{{\beta }_{n}}}} \right\rangle  \\ 
 & \times \left\langle  {{\psi }_{n,{{p}_{y}},-{{\beta }_{n}}}} \right|\hat{V}\left| {{\psi }_{m,{{q}_{y}},{{a}_{m}}}} \right\rangle  \\ 
\end{aligned} \right\}\left[ \frac{-{{e}^{-\frac{\lambda _{m}^{2}}{{{M}^{2}}}}}}{\lambda _{m}^{2}-\lambda _{n}^{2}}-\frac{{{M}^{2}}\left( {{e}^{-\frac{\lambda _{m}^{2}}{{{M}^{2}}}}}-{{e}^{-\frac{\lambda _{n}^{2}}{{{M}^{2}}}}} \right)}{{{\left( \lambda _{m}^{2}-\lambda _{n}^{2} \right)}^{2}}} \right]}} \right\}.
\ee		
Use this result and Eq. (\ref{eqn:eq44}) to obtain
\be 
{{C}_{J2b}}={{C}_{J2bA}}+{{C}_{J2bB}}=\frac{{{L}_{z}}}{2\pi V}\left\{ \sum\limits_{n,{{p}_{y}},{{\beta }_{n}}}{\sum\limits_{m,{{q}_{y}},{{\alpha }_{m}}}{\left\{ \begin{aligned}
  & \left\langle  {{\psi }_{n,{{p}_{y}},{{\beta }_{n}}}} \right|{{\sigma }_{z}}\left| {{\psi }_{n,{{p}_{y}},-{{\beta }_{n}}}} \right\rangle  \\ 
 & \times \left\langle  {{\psi }_{n,{{p}_{y}},-{{\beta }_{n}}}} \right|\hat{V}\left| {{\psi }_{m,{{q}_{y}},{{\alpha }_{m}}}} \right\rangle  \\ 
 & \times \left\langle  {{\psi }_{m,{{q}_{y}},{{\alpha }_{m}}}} \right|\hat{V}\left| {{\psi }_{n,{{p}_{y}},{{\beta }_{n}}}} \right\rangle  \\ 
\end{aligned} \right\}\left[ \frac{\left( {{e}^{-\frac{\lambda _{n}^{2}}{{{M}^{2}}}}}-{{e}^{-\frac{\lambda _{m}^{2}}{{{M}^{2}}}}} \right)}{\lambda _{m}^{2}-\lambda _{n}^{2}} \right]}} \right\}.
\ee		
Adding to this the expression of ${{C}_{J2a}}+{{C}_{J1}}$ given in Eq. (\ref{eqn:eq40}), we finally conclude that
\be 
{{C}_{J1}}+{{C}_{J2}}=\left( {{C}_{J1}}+{{C}_{J2a}} \right)+{{C}_{J2b}}=0.
\ee	
Therefore we have shown that ${{C}_{J}}$ is independent of $\hat{V}$ up to the order ${{\hat{V}}^{2}}$ for arbitrary $M>0$.

\end{widetext}

\end{document}